\documentclass[prl,noshowpacs,english,superscriptaddress,twocolumn]{revtex4}
\usepackage{amsmath,amssymb,amsfonts,mathrsfs}
\usepackage{amsthm}
\usepackage{CJK}
\usepackage{bm}
\usepackage{babel}
\usepackage{graphicx}
\allowdisplaybreaks
\usepackage{braket}
\usepackage[breaklinks]{hyperref}
\usepackage{color}
\hypersetup{colorlinks=true, linkcolor=blue, citecolor=blue, filecolor=blue, urlcolor=blue}
\usepackage{multirow}
\UseRawInputEncoding

\begin{document}

\title{Straight nodal-line phonons in symmorphic space groups}

\author{Guang Liu}
\affiliation{Department of Physics $\&$ Institute for Quantum Science and Engineering, Southern University of Science and Technology,
Shenzhen 518055, P. R. China.}
\affiliation{Guangdong Provincial Key Laboratory of Computational Science and Material Design, Southern University of Science and Technology.}

\author{Yuanjun Jin}
\affiliation{Department of Physics $\&$ Institute for Quantum Science and Engineering, Southern University of Science and Technology,
Shenzhen 518055, P. R. China.}
\affiliation{Guangdong Provincial Key Laboratory of Computational Science and Material Design, Southern University of Science and Technology.}
\author{Zhongjia Chen}
\affiliation{Department of Physics $\&$ Institute for Quantum Science and Engineering, Southern University of Science and Technology,
Shenzhen 518055, P. R. China.}
\affiliation{Guangdong Provincial Key Laboratory of Computational Science and Material Design, Southern University of Science and Technology.}
\affiliation{Department of Physics, South China University of Technology, Guangzhou 510640, P. R. China}

\author{Hu Xu}
\email{xuh@sustech.edu.cn}
\affiliation{Department of Physics $\&$ Institute for Quantum Science and Engineering, Southern University of Science and Technology,
Shenzhen 518055, P. R. China.}
\affiliation{Guangdong Provincial Key Laboratory of Computational Science and Material Design, Southern University of Science and Technology.}

\begin{abstract}
Based on first-principles calculations and effective model analysis, all possible straight nodal-line phonons in symmorphic space groups are uncovered. A classification of two-band $\emph{\textbf{k}}\cdot \emph{\textbf{p}}$ models at an arbitrary point on the nodal lines is carried out, and these nodal-line phonons are found to locate along the high symmetry lines of the Brillouin zone. According to our classification, there are two types of nodal-line phonons in symmorphic space groups. The first type possesses a linear dispersion perpendicular to the nodal lines, while the second type exhibits a quadratic dispersion. The underlying mechanisms are revealed by symmetry analysis, and the types of nodal-line phonons can be distinguished through the little group of the high symmetry lines. Our work not only classifies nodal-line phonons but also provides guidance to search for topological nodal lines in bosonic systems.
\end{abstract}


%Use showkeys class option if keyword %display desired

\maketitle

\makeatletter
\def\@hangfrom@section#1#2#3{\@hangfrom{#1#2#3}}
\makeatother

%Condensed matter system with intrinsic topological quantum state is one of the most important progress of physics research\cite{RevModPhys.82.3045,RevModPhys.83.1057,RevModPhys.90.015001}. The emerging physics such as Quantum spin Hall (QSH) effect, the quantum anomalous Hall(QAH) effect, magnetic monopole, axion, and Majorana fermion which fundamentally changed people's understanding of electronic states\cite{weng2015quantum,PhysRevLett.95.226801,PhysRevLett.95.146802,PhysRevLett.112.017205,Nayake1501870,qi2009inducing,PhysRevLett.122.206401,PhysRevLett.100.096407}. In terms $of physical properties, these novel topological quantum states are mainly derived from the nontrivial energy band degeneracies. According to the dimensionality of the energy band degeneracy in the Brillouin zone (BZ), these peculiar degeneracy can be divided into zero-dimensional nodes\cite{PhysRevB.85.195320,PhysRevB.83.205101,weng2015weyl}, one-dimensional nodal rings and ines\cite{PhysRevB.99.075130,PhysRevB.99.121106,PhysRevB.93.121113,PhysRevB.92.081201,hirayama2017topological,yan2017nodal,PhysRevB.93.035138,deng2019nodal}, and two-dimensional nodal surfaces.\cite{C6NR00882H,PhysRevB.97.115125,PhysRevB.93.085427}

\subsection{INTRODUCTION}
Topological quantum states have attracted much attention in current condensed-matter physics and materials physics \cite{RevModPhys.82.3045,RevModPhys.83.1057,RevModPhys.90.015001}. Topological semimetal is a new class of quantum materials that possesses stable band crossings near the Fermi energy, resulting in many promising properties, such as negative magnetoresistance, chiral magnetic effect, anomalous Hall effect. According to the dimension of the band nodes in the Brillouin zone (BZ), these peculiar crossings can be divided into zero-dimensional nodes\cite{PhysRevB.85.195320,PhysRevB.83.205101,weng2015weyl}, one-dimensional nodal rings and lines\cite{PhysRevB.99.075130,PhysRevB.99.121106,PhysRevB.93.121113,PhysRevB.92.081201,hirayama2017topological,yan2017nodal,PhysRevB.93.035138,deng2019nodal}, and two-dimensional nodal surfaces\cite{C6NR00882H,PhysRevB.97.115125,PhysRevB.93.085427}. Topological nodal-line (NL) semimetal has various ways of band touching along one-dimensional curves in momentum space. The classification of these nodal lines in electronic systems are so far mainly based on band energy degeneracy, band dispersion, and nodal line shape. However, electrons are fermions, which are limited by the Pauli exclusion principle, leading to limited NLs near the Fermi level in electron systems.

 %Only electrons located near the Fermi level contribute to electrical transport which means the nodal lines we can find in the electronic energy band are very limited.

%Nodal-line semimetals expand topological materials beyond topological insulators and Weyl semimetals\cite{PhysRevB.92.081201,C6NR00882H}, which have extended band touching along one dimensional curves in $k$ space and protected by symmetry of space inversion. Recently, Yu Zhiming et.al. have classified the nodal lines in electronic systems with spin-orbit coupling (SOC) based on order of the dispersion relationship on the plane transverse to the nodal line\cite{PhysRevB.99.121106}. But electrons are fermions, which are limited by Pauli's exclusion principle. Only electrons located near the Fermi level contribute to electrical transport which means the high-order dispersion nodal lines we can find in the electronic energy band are very limited.

Besides electrons, phonons are another important elementary excitations in condensed-matter physics. The concept of topology has recently been introduced into phonon systems, greatly enriching the physics of symmetry-protected topological states\cite{PhysRevB.101.085202,PhysRevLett.115.104302,PhysRevB.99.174306,
PhysRevB.96.064106,PhysRevB.98.220103,Susstrunk47,PhysRevB.97.035442,PhysRevLett.117.068001,liu2018berry,susstrunk2015observation,mousavi2015topologically,he2016acoustic}. Topological phonons have potential applications in electron-phonon coupling, multiphonon process, and thermal transports. Besides, phonons are bosons, so the entire frequency range of the phonon spectrum can be physically observed, which can provide us with more material samples to improve the classification of topological NLs. Very recently, nontrivial topological NL phonons in crystalline solids have been studied, including helical NLs in MoB$_2$\cite{PhysRevLett.123.245302}, light operated NL phonons in oxide perovskites\cite{peng2020topological}, and Weyl nodal straight line phonons in MgB$_2$\cite{PhysRevB.101.024301}. However, these works focus only on the research of the NLs in a certain kind of material. There still lacks an extensive classification of the NLs in the phonon systems.

%Therefore, the possibility of finding these topological nodal lines is more abundant.

\linespread{1.1}
\begin{table*}
\centering
\setlength{\tabcolsep}{5mm}
\renewcommand{\tablename}{Table.}
\caption{Linear nodal lines and quadratic nodal lines in symmorphic space groups and their corresponding symmetries.}
\begin{tabular}{|c|c|l|c|c|l|}
\hline
\ line type & symmetry& location&Nodal line & symmetry& location\\
\hline
\multirow{28}{*}{Linear} &\multirow{8}{*}{$C_3+\mathcal{PT}$} &147 (0,0,w)  & \multirow{29}{*}{Quadratic} &\multirow{10}{*}{$C_{2v}+\mathcal{S}_4\mathcal{T}$} &111	(0,0,w)\\
 &~&                                                       147 (1/3,1/3,w)  &~&~&                                                                          111  (1/2,1/2,w)\\
 &~&                                                         148   (w,w,w)  &~&~&  115	(1/2,1/2,w)  \\
 &~&                                                      164 (1/3 ,1/3,w)  &~&~&  115	    (0,0,w)  \\
 &~&                                                      175  (1/3,1/3,w)  &~&~&  119	(w,w,-w)  \\
 &~&                                                          200  (w,w,w)  &~&~&  121	(w,w,-w)  \\
 &~&                                                          202  (w,w,w)  &~&~&  215	(1/2,1/2,w)  \\
 &~&                                                          204  (w,w,w)  &~&~&  215	(0,w,0)  \\
 \cline{2-3}
                        &\multirow{20}{*}{$C_{3v}$}&          156  (0,0,w)  &~&~&  216	(w,w,0)  \\
&~&                                                           157  (0,0,w)  &~&~&  217	(w,-w,w)  \\
\cline{5-6}
&~&                                                       157  (1/3,1/3,w)  &~&$C_3+M_z\mathcal{T}$&  174	(0,0,w)\\
\cline{5-6}
&~&                                                     157  (-1/3,-1/3,w)  &~&\multirow{2}{*}{$C_{3v}+M_z\mathcal{T}$}&  187	(0,0,w)\\
&~&  160	(w,w,w)                                                         &~&~&  189	(0,0,w)\\
\cline{5-6}
&~&  162	(0,0,w)                                                         &~&\multirow{3}{*}{$C_4+\mathcal{PT}$}&  83	(0,0,w)\\
&~&  162	(1/3,1/3,w)                                                     &~&~&  83	(1/2,1/2,w)\\
&~&  164	(0,0,w)                                                         &~&~&  87	(w,w,-w)\\
\cline{5-6}
&~&  166    (0,0,w)                                                         &~&\multirow{10}{*}{$C_{4v}$}&  99	(0,0,w)\\
&~&  183	(1/3,1/3,w)  &~&~&  99	(1/2,1/2,w)\\
&~&  189	(1/3,1/3,w)  &~&~&  107	(w,w,-w)\\
&~&  189	(-1/3,-1/3,w) &~&~&  123	(0,0,w)\\
&~&  191	(1/3,1/3,w)  &~&~&  123	(1/2,1/2,w)\\
&~&  215	(w,w,w)      &~&~&  139	(w,w,-w)\\
&~&  216	(w,w,w)      &~&~&  221	(1/2,1/2,w)\\
&~&  217	(w,w,w)      &~&~&  221	(0,w,0)\\
&~&  221	(w,w,w)       &~&~&  225	(w,w,0)\\
&~&  225	(w,w,w)       &~&~&  229	(w,-w,w)\\
\cline{5-6}
&~&  229    (w,w,w)       &~&$C_6+\mathcal{PT}$&  175	(0,0,w)\\
\cline{5-6}
&~&  ~       &~&\multirow{2}{*}{$C_{6v}$}& 183	(0,0,w)\\
\cline{1-3}
 \multirow{1}{*}{Quadratic}&$C_{2}+\mathcal{S}_4\mathcal{T}$&  81	(0,0,w) &~&~& 191 (0,0,w)\\ \cline{2-3}
\hline
\end{tabular}
\end{table*}

In this work, the NL phonons are classified according to the order of the phonon band dispersion relation in the plane perpendicular to
the NL. To explore this, a systematic symmetry analysis of the high-order NLs in 73 symmorphic space groups is performed, and all possible straight NLs along the high symmetry lines are listed in Table I. It is found that the linear nodal line(LNL) is protected by the point group symmetry $C_3+\mathcal{PT}$ or $C_{3v}$. Quadratic nodal line(QNL) is protected by one of eight symmetries: $C_{2}+\mathcal{S}_4\mathcal{T},C_{2v}+\mathcal{S}_4\mathcal{T},C_3+M_z\mathcal{T},C_{3v}+M_z\mathcal{T},C_4+\mathcal{PT},C_{4v},C_6+\mathcal{PT}$, and $C_{6v}$. It should be noted that the above-mentioned symmetries in the form $A+B$ represent a point group symmetry plus an operation, i.e., the former represents the little group corresponding to the nodal line, and the latter represents the symmetry operation that needs to be satisfied after considering the time reversal symmetry ($\mathcal{T}$).

%There we classify the nodal lines according to the order of the phonon band dispersion relationship. Then a series of fundamental questions naturally arise: \emph{What is the highest order allowed? What additional symmetry does it need? What are their topological properties?} Inspired by these problems and challenges, here we propose a systematic investigation of the higher order weyl nodal lines of a three-dimensional phonon system. We find the stable high-order nodal lines along the high symmetry line of the BZ by symmetry analysis. We can search the  space group corresponding to the different crystal from 73 symmorphic space group. We list all the space group correspond to the nodal line with different order terms in the Table I. The key findings are as follows:(i)We find that beyond the linear weyl modal line, QNL  is the only stable possibilities,  that is, there are no symmetry-protected nodal line with leading order dispersion (along any direction) higher than the second order. (ii)We find all the existing nodal lines in the symmorphic space group and their corresponding point group symmetry protection.  (iii)The nodal lines that protected by point group $c_3+\mathcal{PT}$ and $c_{3v}$ are linear NLs. Considering an additional $M_zT$ symmetry, the first-order term will be eliminated and the nodal line becomes QNL.

\subsection{RESULTS AND DISCUSSION}
To understand the properties of the NL phonons, we first show the nodal-line solution in terms of the two-band
$\emph{\textbf{k}}\cdot \emph{\textbf{p}}$ effective Hamiltonian. Generally, the crossings of two phonon branches can be described by a $2\times2$ effective Hamiltonian, namely

\vspace{-2.5 ex}

\begin{equation}
H(\emph{\textbf{q}})=d(\emph{\textbf{q}})\sigma_++d^*(\emph{\textbf{q}})\sigma_-+f(\emph{\textbf{q}})\sigma_z
\end{equation}
where $H$ is referenced to the frequency of an arbitrary point on the NL phonon, $d(\emph{\textbf{q}})$ represents a complex function, $f(\emph{\textbf{q}})$ represents a real function,  $\emph{\textbf{q}}$ denotes the wave vector which is restricted to the plane perpendicular to  the NL, $\sigma_\pm=\sigma_x\pm i\sigma_y$, and $\sigma_{0,x,y,z}$ are the Pauli matrices. The leading order of $d(\emph{\textbf{q}})$ and $f(\emph{\textbf{q}})$ determine the classification of NLs. If the leading orders of $d(\emph{\textbf{q}})$ and $f(\emph{\textbf{q}})$ are linear, the nodal line can be named as LNL. By analogy, we can define the QNL.

%Through high-throughput search and kp analysis, we have found all the stable straight nodal lines in the symmorphic space group. They are mainly divided into linear nodal and quadratic nodal line.

%Through symmetry analysis, some materials belonging to different space groups which can host NLs have been discovered. Here, we identify two material examples that host the kinds of NLs discovered in this work.

\begin{figure}
     \centering
     \renewcommand{\figurename}{FIG.}
     \includegraphics[scale=0.4]{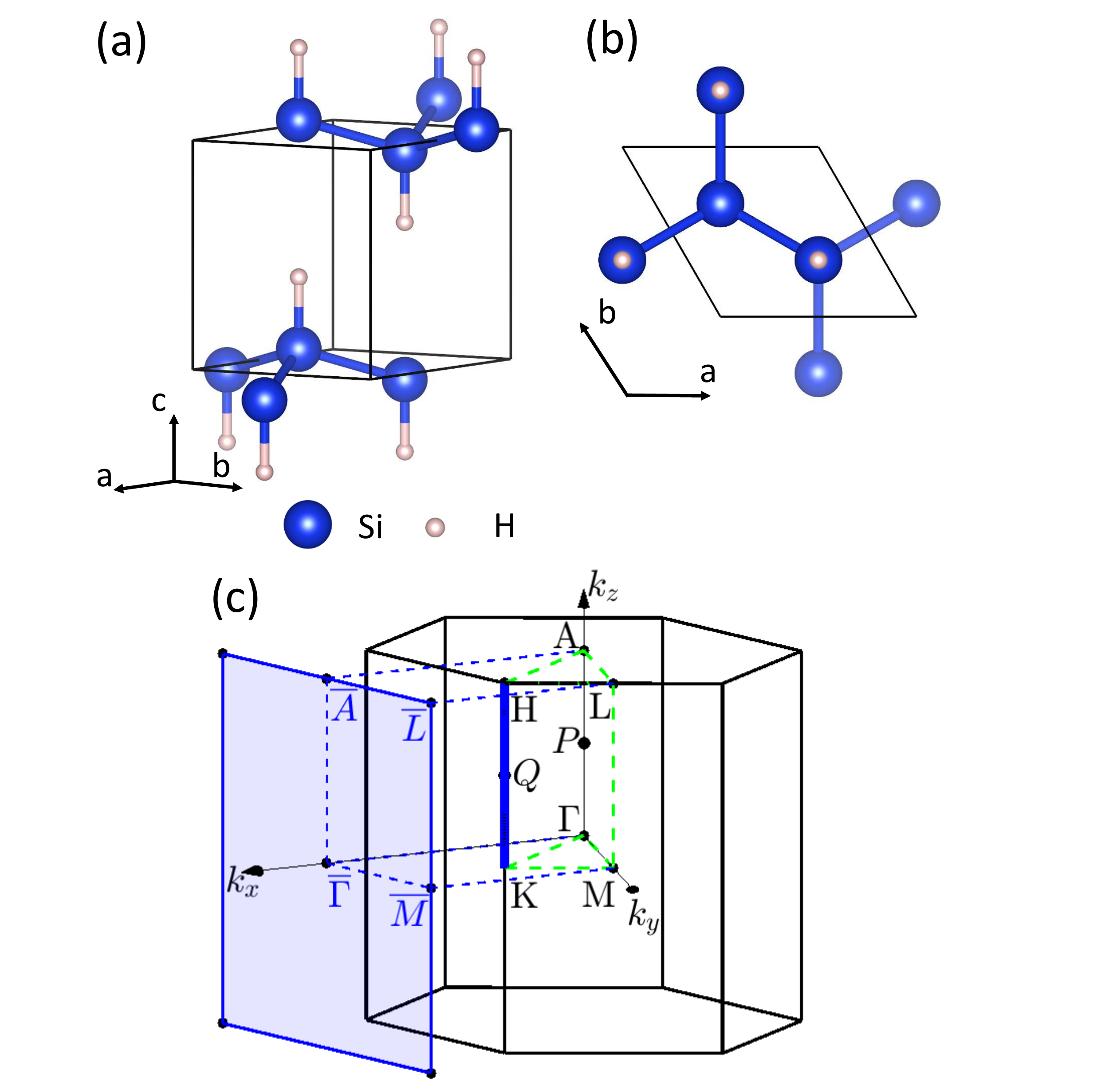}
     \caption{Crystal structure and the corresponding BZ. (a) Side and (b) top views of SiH crystallized in a hexagonal structure with space group $P\overline{3}m1$(No.164). (c) The bulk BZ and its corresponding surface BZ projected on the (100) surface.}
\end{figure}

\subsubsection{Linear nodal line}
As listed in Table I, the LNL phonons are stabilized by symmetry $C_{3v}$ or $C_3+\mathcal{PT}$.
For little group $C_{3v}$, there is a two-dimensional irreducible representation corresponding to a double degenerate NL. An arbitrary point on this path is invariant under the little group which contains two generators: the threefold rotation $C_{3z}$ and vertical mirror $M_x$. The relevant representation is $\Gamma_3$, of which the basis functions are $\{ (\textbf{\emph{S}}_x-i\textbf{\emph{S}}_y),-(\textbf{\emph{S}}_x+i\textbf{\emph{S}}_y)\}$. The matrix representations of the generators can be expressed in the above basis as

\vspace{-2.5 ex}

\begin{equation}
D(C_{3z})=e^{2i\sigma_z\pi/3},
D(M_x)=\sigma_x.
\end{equation}

The Hamiltonian is constrained by the symmetry transformations, namely

\vspace{-2 ex}

\begin{equation}
D(R)H_{eff}(\emph{\textbf{q}})D^{-1}(R)=H_{eff}(R\emph{\textbf{q}}),
\end{equation}
where $R$ denotes the corresponding operator $C_{3z}$ and $M_x$,  which acts on $\emph{\textbf{q}}$. The constraint Eq. (3) gives

\vspace{-2 ex}

\begin{equation}
\begin{aligned}
f(q_+,q_-)&=f(e^{i2\pi/3}q_+,e^{-i2\pi/3}q_-),\\
e^{i4\pi/3}d(q_+,q_-)&=d(e^{i2\pi/3}q_+,e^{-i2\pi/3}q_-),\\
-f(q_+,q_-)&=f(-q_-,-q_+),\\
d^*(q_+,q_-)&=d(-q_-,-q_+),
\end{aligned}
\end{equation}
where $q\pm=q_x\pm iq_y$. Then we expand Eq. (4)
and remain the lowest orders as

\vspace{-2 ex}

\begin{equation}
\begin{aligned}
f(\emph{\textbf{q}})&=a_1(q_+^3+q_-^3),\\
d(\emph{\textbf{q}})&=ia_2q_-,
\end{aligned}
\end{equation}
where $a_{1}$ and $a_{2}$ are real parameters. Thus, the corresponding effective Hamiltonian retained to the leading
order can be simplified to

\begin{equation}
H_{eff}(\textbf{\emph{q}})=iaq_-\sigma_++H.c.,
\end{equation}
where $a$ is a real parameter, and Eq. (6) indicates that the dispersion relation in the plane perpendicular to the NL is linear. According to the definition, the doubly degenerate nodal lines that are protected by $C_{3v}$  are LNL.

%where $R$ denotes the corresponding operator $C_{3z}$ and $M_x$  which are acting on $\emph{\textbf{q}}$. We carry out a Taylor expansion on the $f(\emph{\textbf{q}})$ in Eq.(1) and only keep the third term. This is because the linear term and quadratic term in $f(\emph{\textbf{q}})$ are eliminated by the constraint of $D(C_{3z})H_{eff}(\textbf{\emph{q}})D^{-1}(C_{3z})=H_{eff}(C_{3z}\emph{\textbf{q}})$, that is, $\sigma_z$ is invariant under $C_{3z}$[$D(C_{3z})\sigma_zD(C_{3z})^{-1}= \sigma_z$], but $R_{3z}$ rotates $\emph{\textbf{q}}$. The constant term of $f(\emph{\textbf{q}})$ is excluded  because of $D(M_x)\sigma_zD(M_x)^{-1}=-\sigma_z$. Similarly, for the Taylor expansion of $d(\emph{\textbf{q}})$, the zero order term cannot be retained because it is not invariant under the operation of $C_{3z}$, namely,  $D(C_{3z})\sigma_\pm D(C_{3z})^{-1}= e^{\pm i4¦Ð/3}\sigma_\pm$. Under the constraint of the generator $C_{3z}$ in crystal symmetry, the linear term in $d(\emph{\textbf{q}})$ must take the form of $cq_-\sigma_++c.c.$, where $q\pm=q_x\pm iq_y$ and $c$ is a constant. Thus, the corresponding effective Hamiltonian retained to the leading order can be simplified to

%\vspace{-3.5 ex}

%\begin{equation}
%H_{eff}(\emph{\textbf{q}})=\alpha q_-\sigma_++H.c.,
%\end{equation}
%where $\alpha$ is a complex parameter, and the Eq.(4) indicates that the dispersion relation in the plane perpendicular to the NL is linear. According to the definition, we conclude that the doubly degenerate nodal lines protected by $C_{3v}$  are LNL.

For little group $C_3$, the $\mathcal{PT}$ symmetry satisfies the commutation relationship $[\mathcal{PT},H]$. We assume that there is a set of basis $\{\psi, \mathcal{PT}\psi\}$ which are the eigenstates with common eigenvalue of Hamiltonian. Selecting a set of basis functions $\{-i(x+iy),i(x-iy)\}$ belong to two one-dimensional irreducible representations of $C_3$, which meets $\psi_2=\mathcal{PT}\psi_1$($\{\psi_1,\psi_2\}$). So there is an essential degenerate NL. An arbitrary point on this path is invariant under the threefold rotation $C_{3z}$ and $\mathcal{PT}$. The relevant representations are $\{\Gamma_2,\Gamma_3\}$, of which the basis are $\{-i(x+iy),i(x-iy)\}$ . The matrix representations of the operators can be expressed in the above basis as
%$\{-(\textbf{\emph{S}}_x+i\textbf{\emph{S}}_y),(\textbf{\emph{S}}_x-i\textbf{\emph{S}}_y)\}$ or

\vspace{-2 ex}

\begin{equation}
D(C_{3z})=e^{2i\sigma_z\pi/3},
D(\mathcal{PT})=\sigma_x\mathcal{K},
\end{equation}
where $\mathcal{K}$ denotes the complex conjugation operator. The corresponding effective Hamiltonian reads

\vspace{-3 ex}

\begin{equation}
H_{eff}(\textbf{\emph{q}})=\alpha q_-\sigma_++H.c.,
\end{equation}
where $\alpha$ is a complex parameter. As given in Eq.(8), the nodal lines protected by $C_3+\mathcal{PT}$ are also LNL.

%Through the guidance of model analysis, we found an
According to our model analysis, we find that SiH hosts the LNL phonons. As illustrated in Figs. 1(a) and 1(b), SiH  crystallizes in a trigonal lattice with symmorphic space group $P\overline{3}m1$ (No. 164). The optimized lattice constants are $a=3.861$ {\AA}  and $c=4.804$ {\AA}. The hexagonal bulk BZ and its corresponding (100) surface BZ are shown in Fig. 1(c). There is a nodal line along the high symmetry line $K$-$H$ [see blue bold line in Fig. 1(c)].

%$HSi$ is a bulk material with $D_{3d}$ crystal symmetry, which is in good agreement with our analysis. As illustrated in Figs. 1(a) and 1(b), the  $HSi$  crystallizes in a trigonal lattice with a space group $P\overline{3}m1$ (No.164). The optimized lattice constants are $a=3.861 {\AA} $ and $c=4.804 {\AA} $. The hexagonal bulk BZ and its corresponding (100) and (001) surface BZs are shown in Figs. 1(c).

As shown in Fig. 2(a), the phonon spectrum of SiH is plotted. It is noteworthy that we can find a doubly degenerate phonon
band along the high symmetry line $K$-$H$ near 19 THz, which is contributed from the 8th and 9th phonon branches. These two crossing bands belong to states which have opposite eigenvalues $e^{¡À2/3i\pi}$ of rotation symmetry operation $C_{3z}$, which protects the NL along $K$-$H$. Because the symmetries of the phonon momentum obey a $C_{3v}$ point group along the $K$-$H$ direction, the doubly degenerate lines belong to a two-dimensional irreducible representation. The three-dimensional(3D) representation of two-dimensional phonon dispersion at any point on the nodal line is plotted in Fig. 2(b). There is a crossing of the two branches in the shape of a Dirac cone at the Q point in momentum space, which indicates that the dispersion relation is linear. These results from first-principles calculations agree well with our symmetry analysis. The band dispersion around a generic point also shows an obvious linear relationship in Fig. 2(c), indicating the existence of the LNL.

In Fig. 2(c), we plot the calculated phonon local density of states (LDOS) projected on the semi-infinite
(100) surface of SiH. The topological drumhead surface states are clearly visible, which means that it is possible to observe surface states experimentally. As expected, the surface states begin at a crossing point on a high-symmetry line $\overline\Gamma-\overline M$ and terminate at a crossing point on another high symmetry line $\overline A-\overline L$. These results are consistent with our theoretical analysis and prove that the surface states come from the LNL phonons.

\begin{figure}
\centering
\renewcommand{\figurename}{FIG.}
\setlength{\abovecaptionskip}{-0.2 cm}
\includegraphics[scale=0.38]{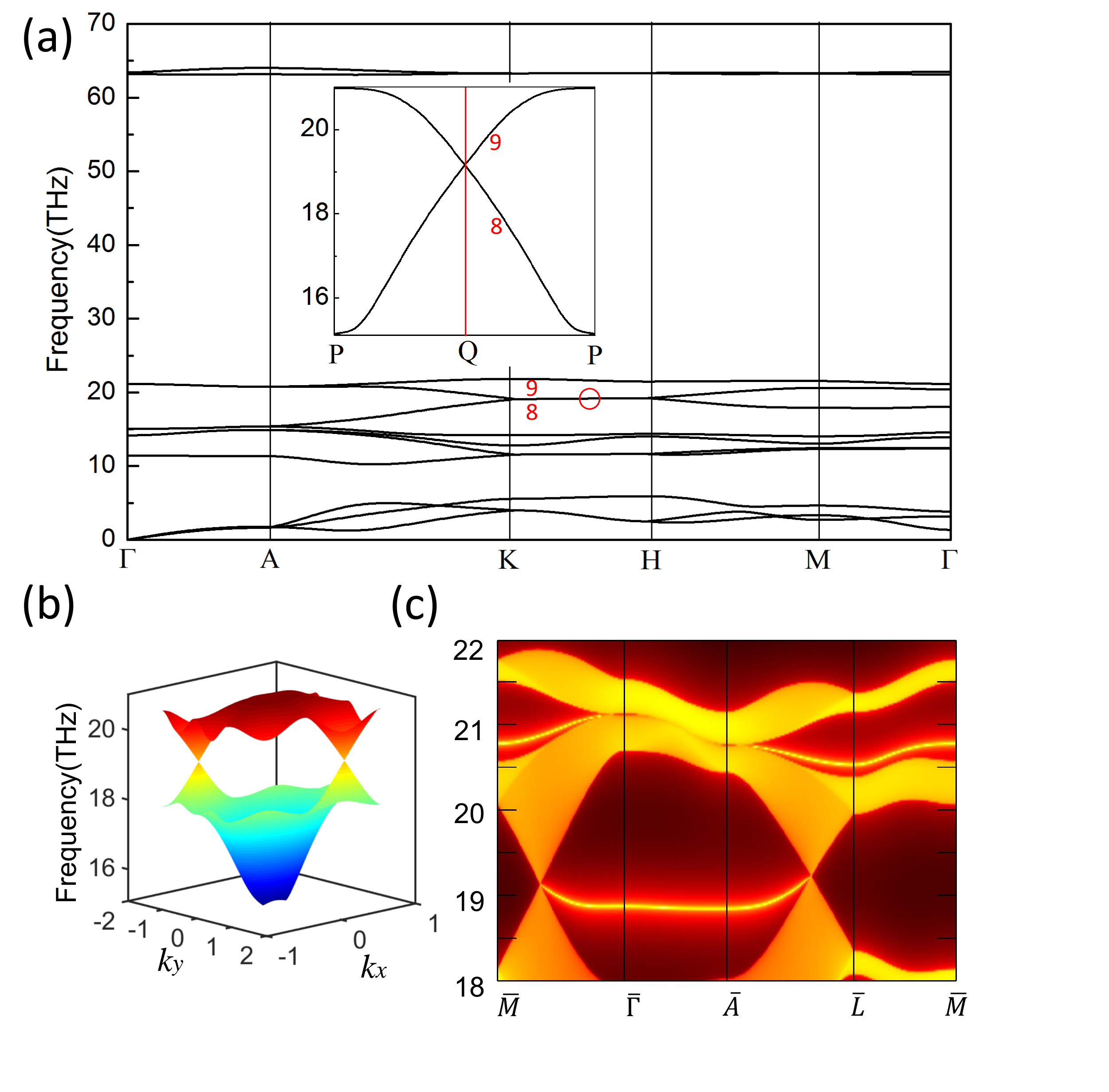}
\caption{(a) The phonon spectrum of SiH along the high-symmetry lines. P and Q are the midpoints of high symmetry lines $\Gamma$-$A$ and $K$-$H$, respectively. (b) The 3D representation of the LNL phonons at any point on the nodal line. Here, we visualize them by selecting a node on the $K$-$H$ at $k_z$=0.2. (c) LDOS projected on the semi-infinite (100) surface. Yellow areas denote the projections of bulk phonon branches, and yellow lines represent the nontrivial phonon surface states.}
\end{figure}

\subsubsection{Quadratic nodal line}
Our symmetry analysis shows that the QNL phonons are protected by eight different types of symmetries, including $C_{2}+\mathcal{S}_4\mathcal{T},C_{2v}+\mathcal{S}_4\mathcal{T},C_3+M_z\mathcal{T},C_{3v}+M_z\mathcal{T},C_4+\mathcal{PT},C_{4v},C_6+\mathcal{PT}$, and $C_{6v}$, respectively. Here, we take $C_{2v}+\mathcal{S}_4\mathcal{T}$ as an example, and more detailed results are provided in the supplementary materials (SM)\cite{SM}.

Momentum with the $C_{2v}$ symmetry on the line can be invariant under $\mathcal{S}_4\mathcal{T}$. So we can drive that $[\mathcal{S}_4\mathcal{T}, H]$. We consider two states $\{\psi, \mathcal{S}_4\mathcal{T}\psi\}$, which are the eigenstates with the common eigenvalue of Hamiltonian. In addition, $C_{2v}$ has two one-dimensional irreducible representations that are complex, and the relevant representations are $\{\Gamma_3,\Gamma_4\}$, of which the basis is $\{ x,y\}$.  Considering the operation $\mathcal{S}_4\mathcal{T}$, two one-dimensional irreducible representations are combined into a two-dimensional irreducible representation. So there is a doubly degenerate band. The little group contains two generators: the twofold rotation $C_{2z}$ and the vertical mirror $M_x$. The matrix representations of the generators can be expressed on the above basis as

\vspace{-2.5 ex}

\begin{equation}
D(C_{2z})=-\sigma_0, D(M_x)=\sigma_z.
\end{equation}

Considering two operators $C_{3z}$ and $M_x$  which are acting on $\emph{\textbf{q}}$, the constraint Eq. (3) gives

\vspace{-2 ex}

\begin{equation}
\begin{aligned}
f(q_+,q_-)&=f(-q_+,-q_-),\\
d(q_+,q_-)&=d(-q_+,-q_-),\\
f(q_+,q_-)&=f(-q_-,-q_+),\\
-d(q_+,q_-)&=d(-q_-,-q_+).
\end{aligned}
\end{equation}

Then we expand Eq. (10) and remain the lowest orders as

\vspace{-2 ex}

\begin{equation}
\begin{aligned}
f(\emph{\textbf{q}})&=a_1(q_+^2+q_-^2)+a_2q_+q_-,\\
d(\emph{\textbf{q}})&=\alpha(q_+^2-q_-^2).
\end{aligned}
\end{equation}

Thus, the corresponding effective Hamiltonian retained to the leading
order can be simplified to
\begin{equation}
\begin{aligned}
H_{eff}&(\textbf{\emph{q}})=[a_1(q_+^2+q_-^2)+a_2q_+q_-]\sigma_z\\
&+\alpha(q_+^2-q_-^2)\sigma_++H.c.,
\end{aligned}
\end{equation}
which indicates that the dispersion relation in the plane perpendicular to the NL is quadratic. It is natural to conclude that the doubly degenerate phonon NLs protected by $C_{2v}+\mathcal{S}_4\mathcal{T}$ are QNL.

%The Hamiltonian is constrained by these symmetry transformations. Different from LNL effective model, for the Taylor series expansion of $f(\emph{\textbf{q}})$, we see that all the odd-order term are excluded by the constraint of Eq. (3) because of $D(C_{2z})\sigma_zD(C_{2z})^{-1}=\sigma_z$. The leading order in $f(\emph{\textbf{q}})$ is quadratic.  For the Taylor series of $d(\emph{\textbf{q}})$, the zeroth term vanishes because $D(M_{x})\sigma_\pm D(M_{x})^{-1}= -\sigma_\pm$. The odd-order term in $d(\emph{\textbf{q}})$ is also eliminated by the constraint of $D(C_{2z})H_{eff}(\emph{\textbf{q}})D^{-1}(C_{2z})=H_{eff}(C_{2z}\emph{\textbf{q}})$, because $\sigma_x$ is invariant under $C_{2z}$, but $R_{2z}$ rotates q. Due to the constraint of the operator $M_x$, the even term in $d(\emph{\textbf{q}})$ must take the form $cq_xq_y\sigma_++c.c.$. Thus, the corresponding effective Hamiltonian to the leading order can be described as

%\begin{equation}
%H_{eff}(\emph{\textbf{q}})=a(q_x^2+q_y^2)\sigma_z+\alpha q_xq_y\sigma_++H.c.,
%\end{equation}
%where $a$ is a real parameter, and Eq. (8) indicates that the dispersion relation in the plane transverse to the NL is quadratic. It is natural to conclude that the doubly degenerate phonon NLs protected by $C_{2v}+\mathcal{S}_4\mathcal{T}$ are QNL.

\begin{figure}
\centering
\renewcommand{\figurename}{FIG.}
\setlength{\abovecaptionskip}{-0.4 cm}
\includegraphics[scale=0.35]{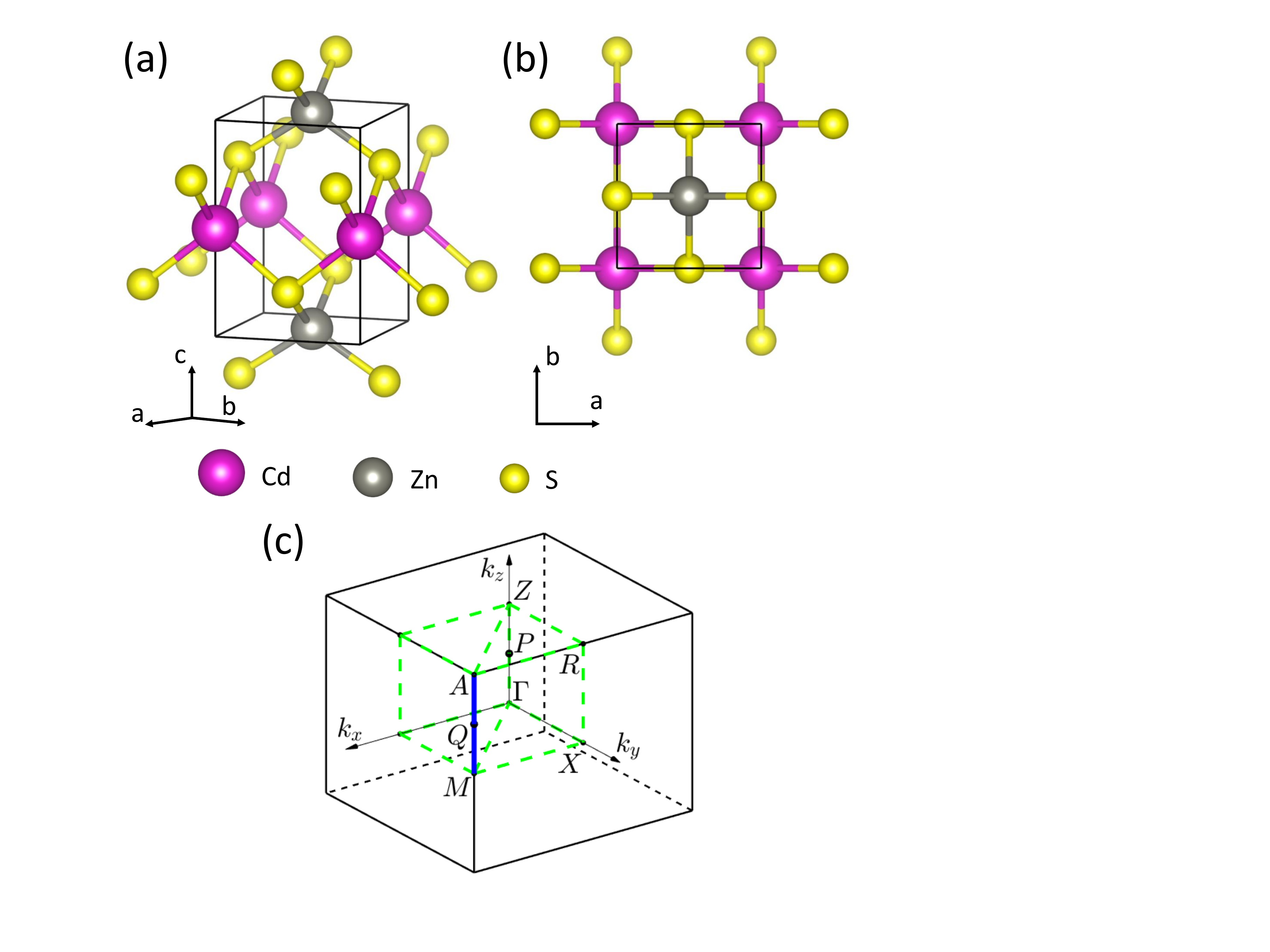}
\caption{Crystal structure and BZ. (a) Side and (b) top views of ZnCdS$_2$ with space group $P\overline{4}m2$ (No. 115). (c) The tetragonal bulk BZ. }
\end{figure}

\begin{figure}
\centering
\renewcommand{\figurename}{FIG.}
\includegraphics[scale=0.43]{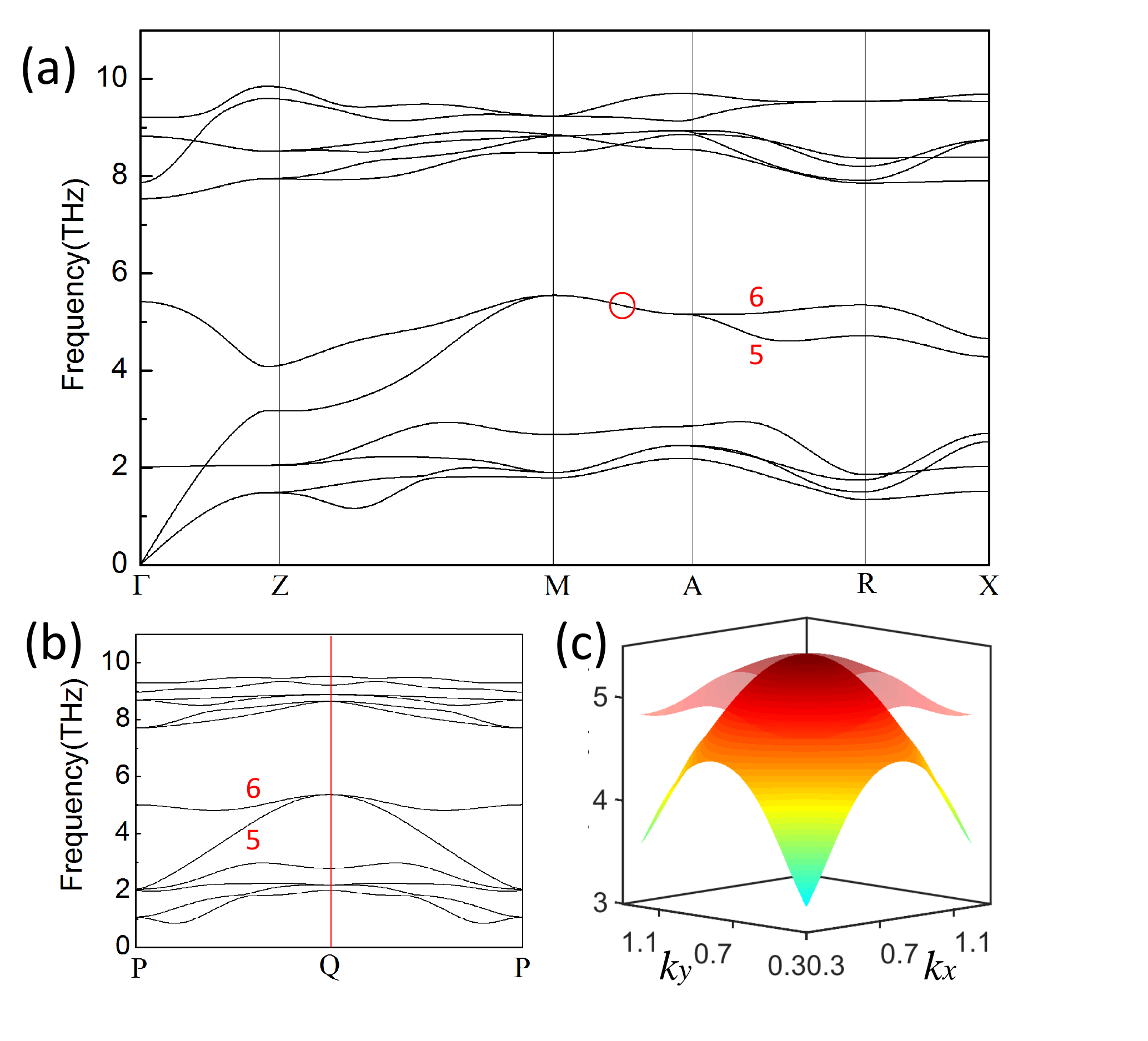}
\caption{(a) The phonon spectrum of ZnCdS$_2$ along the high-symmetry lines. (b) The band dispersion along the path $P$-$Q$-$P$ (in direction perpendicular to the NL). $P$ and $Q$ are in the middle of $\Gamma$-$Z$ and $M$-$A$, respectively. (c)The 3D view of the 5th and 6th bands restricted to the $k_x$-$k_y$ plane at any point on the nodal line. Here, we visualize them by selecting a node on the $M$-$A$ at $k_z$=0.2.}
\end{figure}

ZnCdS$_2$ is an ideal candidate to possess the QNL phonons. As illustrated in Figs. 3(a) and 3(b), ZnCdS$_2$ crystallizes in a tetragonal lattice with space group $P\overline{4}m2$ (No. 115). The optimized lattice constants are $a=3.930$ {\AA} and $c=5.645$ {\AA}. The tetragonal bulk BZ is shown in Fig. 3(c). We find that there is a NL along the high symmetry line $M$-$A$ with $C_{2v}+\mathcal{S}_4\mathcal{T}$ symmetry.

As shown in Fig. 4(a), the phonon spectrum of ZnCdS$_2$ is plotted. It is noteworthy that we can
find out a doubly degenerate phonon band along the high symmetry line $M$-$A$ near 5 THz, which are
contributed from the 5th and 6th phonon branches [highlighted in Fig. 4(b)]. Because the symmetries of the phonon momentum obey a $C_{2v}$ point group along the $M$-$A$ direction, the doubly degenerate line belongs to two one-dimension irreducible representations. These two crossing bands belong to states that have the opposite
eigenvalues $\pm1$ of mirror reflection symmetry operation $M_x$, respectively, which protects the NL along the high symmetry line $M$-$A$. In Fig. 4(b), we can see an obvious quadratic dispersion relation around the Q point between the 5th and 6th branches. The 3D representation of two-dimensional phonon dispersion at any point on the nodal line is plotted in Fig. 4(c), and it also shows an obvious quadratic relation in the $k_x$-$k_y$ plane. The evidence shows that the NL along $M$-$A$ is indeed a QNL.

To further explore the topological properties of these NLs, the corresponding Berry phase distribution is considered.
In general, the Berry phase of a closed path C in 3D BZ is defined as \cite{rspa.1984.0023}

\vspace{- 3.5 ex}

\begin{equation}
\gamma_C=\oint_C\emph{\textbf{A}}_n(\emph{\textbf{q}})\cdot d\emph{\textbf{q}},
\end{equation}
where $\emph{\textbf{A}}_n(\emph{\textbf{q}})=i\langle \psi_n(\emph{\textbf{q}})|\nabla_\emph{\textbf{q}}|\psi_n(\emph{\textbf{q}})\rangle$ is the Berry connection and
$\psi_n(\emph{\textbf{q}})$ is the Bloch wave function of occupied bands. We select a closed circle
centered at a generic point of the nodal line. Note that the circle lies on the plane perpendicular to the line and cannot cover
another nodal line. Interestingly, we find that the Berry phase for the LNL phonons in SiH along the $K$-$H$ is $\pi$, for QNL in ZnCdS$_2$ along $M$-$A$ is $2\pi$. More details are
included in Table SI of the SM. The results show that the Berry phase are $\pi$ for linear NLs, and $2\pi$ for QNLs.

\vspace{3 ex}

\subsection{CONCLUSIONS}
In summary, using first-principles calculations and effective model analysis, we have systematically studied the symmetry-protected NL phonons in symmorphic space groups. All possible straight NL phonons have been uncovered, and our results show that the highest-order dispersion relation perpendicular to the NL is quadratic. Via symmetry analysis, we have found the existing NL phonons in symmorphic space groups according to the corresponding little groups. We thus offer a feasible way of finding topological NL materials.

%This method can also be applied to the non-symmorphic space group system. By searching for the quotient group that is isomorphic to the point group in the non-symmorphic space group, we can classify the NLs in the non-symmorphic space group.

This work is supported by the National Natural Science
Foundation of China (NSFC, Grants No. 11974160), the Guangdong
Natural Science Funds for Distinguished Young Scholars
(No. 2017B030306008), the fund of the Guangdong Provincial Key Laboratory of Computational Science and Material Design (No.2019B030301001),
and the Center for Computational Science and Engineering at Southern University of Science and Technology.

\newpage
\appendix

\setcounter{figure}{0}
\makeatletter

\makeatother
\renewcommand{\thefigure}{S\arabic{figure}}
\renewcommand{\thetable}{S\Roman{table}}
\renewcommand{\theequation}{S\arabic{equation}}
\begin{center}
	\textbf{
		\large{Supplemental Material for}}
	\vspace{0.2cm}
	
	\textbf{
		\large{
			``Straight nodal-line phonons in symmorphic space groups" }
	}
\end{center}

\vspace{-0.2cm}

\section{Calculation methods}
All the calculations were based on the framework of density functional theory (DFT) \cite{kohn1965self} using the Vienna \emph{ab-initio} Simulation Package (VASP) \cite{PhysRevB.54.11169,kresse1996efficiency}. The generalized gradient approximation (GGA) with the Perdew-Burke-Ernzerhof (PBE) formalism was employed for the exchange-correlation function \cite{PhysRevLett.77.3865,PhysRevLett.78.1396}. The projector augmented wave method was employed to treat core-valence interactions\cite{PhysRevB.59.1758,PhysRevLett.45.566}. For phonon spectra calculations, we used the PHONOPY code to construct the force constants matrices and generate the symmetry information \cite{TOGO20151}. The phonon surface states were calculated using the iterative Green function method \cite{Sancho_1984} with the tight-binding model Hamiltonian carried out by the WannierTools package \cite{WU2018405}.

%To illustrate the topological properties more intuitively, we constructed a phonon tight-binding Hamiltonian with the iterative Green¡¯s function method\cite{Sancho_1984} to calculate the local density of states (LDOS) of phonons by using the WANNIERTOOLS package\cite{WU2018405}.

\section{Details of the $\emph{k} \cdot \emph{p}$ model}

%Herein, we derive the effective Hamiltonian based on the space group using the $\textbf{\emph{k}} \cdot \textbf{\emph{p}}$ model. It is important to note that point group can be treated as a subgroup of the symmorphic space group.  As below, we generate the IRs by performing the REPRES program, which is carried out by a normal-subgroup induction method.

To understand the properties of the phonon NLs, we first show the nodal-line solution in terms of the two-band
$\emph{\textbf{k}}\cdot \emph{\textbf{p}}$ effective Hamiltonian. Generally, the crossings of two phonon branches can be described by a $2\times2$ effective Hamiltonian, namely

\renewcommand{\theequation}{S\arabic{equation}}
\begin{equation}
H(\emph{\textbf{q}})=d(\emph{\textbf{q}})\sigma_++d^*(\emph{\textbf{q}})\sigma_-+f(\emph{\textbf{q}})\sigma_z
\end{equation}
where $H$ is referenced to the frequency of an arbitrary point on the phonon NL, $d(\emph{\textbf{q}})$ represents a complex function, $f(\emph{\textbf{q}})$ represents a real function.

\subsection{Linear nodal lines}

\subsubsection{Point group $C_3$ with $\mathcal{PT}$}
Considering a double degenerate phonon band where the little group is $C_{3}$. We
know that point group $C_{3}$ has two one-dimensional
irreducible representations which are complex conjugate to each other. Considering the symmetry
of combined operator $\mathcal{PT}$, two one-dimensional irreducible representations can be combined into
a two-dimensional irreducible representation. The basis functions are $\{-i(x+iy),i(x-iy)\}$ or $\{-(S_{x}+iS_{y}),(S_{x}-iS_{y})\}$

To construct the $k\cdot p$ effective Hamiltonian around a generic point, we
express the symmetry operators in the basis:

\renewcommand{\theequation}{S\arabic{equation}}
\begin{equation}
\renewcommand{\theequation}{S\arabic{equation}}
\begin{aligned}
&C_{3z}=\left(
\begin{array}
[c]{cc}%
e^{\frac{2i\pi}{3}} & 0\\
0 & e^{\frac{-2i\pi}{3}}%
\end{array}
\right)  ,
C_{3z}^{-1}=\left(
\begin{array}
[c]{cc}%
e^{\frac{-2i\pi}{3}} & 0\\
0 & e^{\frac{2i\pi}{3}}%
\end{array}
\right)  ,\\
&\mathcal{PT}=\left(
\begin{array}
[c]{cc}%
0 & 1\\
1 & 0%
\end{array}
\right) K ,
\mathcal{PT}=\left(
\begin{array}
[c]{cc}%
0 & 1\\
1& 0%
\end{array}
\right) \mathcal{K},
\end{aligned}
\end{equation}
where $\mathcal{K}$ is the complex conjugation operator, and $\sigma_{i}$ are the Pauli
matrices with i =0, x, y, z. The Hamiltonian is required to be invariant under
the symmetry transformations, namely,

\begin{equation}
D(R)H_{eff}(\emph{\textbf{q}})D^{-1}(R)=H_{eff}(R\emph{\textbf{q}}),
\end{equation}
where $R$ denotes the corresponding operator $C_{3z}$ and $\mathcal{PT}$  which are acting on $\emph{\textbf{q}}$. The constraint Eq. (S3) gives

\begin{equation}
\begin{aligned}
f(q_+,q_-)&=f(e^{i2\pi/3}q_+,e^{-i2\pi/3}q_-),\\
e^{i4\pi/3}d(q_+,q_-)&=d(e^{i2\pi/3}q_+,e^{-i2\pi/3}q_-),\\
-f^*(q_+,q_-)&=f(q_+,q_-),\\
d(q_+,q_-)&=d(q_+,q_-).
\end{aligned}
\end{equation}

where $\sigma_{\pm}=(\sigma_{x}\pm i\sigma_{y})/2$. Thus, the corresponding effective Hamiltonian retained to the lowest order of $f(\emph{\textbf{q}})$ and $d(\emph{\textbf{q}})$ can be simplified to
\begin{equation}
H_{eff}=a(q_+^3-q_-^3)\sigma_z+\alpha q_{-}\sigma_{+}+H.c.,
\end{equation}
where $a$ is a real parameter and $\alpha$ is a complex parameter. Thus, we conclude that the
doubly degenerate line at leading order is a linear NL.

\subsubsection{Point group $C_{3v}$}

Considering a double degenerate phonon band where the little group is $C_{3v}$.
We know that point group $C_{3v}$ have a two dimensional irreducible
representation which corresponding to a double degenerate band.
$C_{3v}$ contains two generators: $C_{3z},M_{x}.$ The basis
functions are $\{(S_{x}-iS_{y}),-(S_{x}+iS_{y})\}$

To construct the $k\cdot p$ effective Hamiltonian around a generic point, we
express the symmetry operators in the basis:

\begin{equation}
\begin{aligned}
&C_{3z}=\left(
\begin{array}
[c]{cc}e^{\frac{2i\pi}{3}} & 0\\
0 & e^{\frac{-2i\pi}{3}}
\end{array}
\right)  ,
C_{3z}^{-1}=\left(
\begin{array}
[c]{cc}%
e^{\frac{-2i\pi}{3}} & 0\\
0 & e^{\frac{2i\pi}{3}}%
\end{array}
\right)  ,\\
&M_{x}=\left(
\begin{array}
[c]{cc}%
0 & 1\\
1 & 0
\end{array}
\right)  ,
 M_{x}^{-1}=\left(
\begin{array}
[c]{cc}%
0 & 1\\
1 & 0
\end{array}
\right)  ,
\end{aligned}
\end{equation}

The constraint Eq. (S3) gives

\begin{equation}
\begin{aligned}
f(q_+,q_-)&=f(e^{i2\pi/3}q_+,e^{-i2\pi/3}q_-),\\
e^{i4\pi/3}d(q_+,q_-)&=d(e^{i2\pi/3}q_+,e^{-i2\pi/3}q_-),\\
-f(q_+,q_-)&=f(-q_-,-q_+),\\
d^*(q_+,q_-)&=d(-q_-,-q_+).
\end{aligned}
\end{equation}

Thus, the corresponding effective Hamiltonian retained to the lowest order of $f(\emph{\textbf{q}})$ and $d(\emph{\textbf{q}})$ can be simplified to
\begin{equation}
H_{eff}=a_1(q_+^3+q_-^3)\sigma_z+ia_2q_{-}\sigma_{+}+H.c.,
\end{equation}
where $a_{1,2}$ is a real parameter. Thus, we conclude that the
doubly degenerate line at leading order is a linear NL.

\subsection{Quadratic nodal lines}

\subsubsection{Point group $C_{2v}$ with $S_4\mathcal{T}$}

Considering a double degenerate phonon band where the little group is $C_{2v}$.
We know that point group $C_{2v}$ has two one-dimensional
irreducible representations which are complex conjugate to each other. Considering the symmetry
operator $S_4\mathcal{T}$, two one-dimensional irreducible representations can be combined into
a two-dimensional irreducible representation. So there is a
double degenerate band. $C_{2v}$ contains two generators:
$C_{2z},M_{x}$. The basis functions are $\{x,y\}$.

To construct the $k\cdot p$ effective Hamiltonian around a generic point, we
express the symmetry operators in the basis:

\begin{equation}
\begin{aligned}
&C_{2z}=\left(
\begin{array}
[c]{cc}%
-1 & 0\\
0 & -1
\end{array}
\right)  ,
C_{2z}^{-1}=\left(
\begin{array}
[c]{cc}%
-1 & 0\\
0 & -1
\end{array}
\right)  ,\\
&M_{x}=\left(
\begin{array}
[c]{cc}%
1 & 0\\
0 & -1
\end{array}
\right)  ,M_{x}^{-1}=\left(
\begin{array}
[c]{cc}%
1 & 0\\
0 & -1
\end{array}
\right)  ,\\
\end{aligned}
\end{equation}

The constraint Eq. (S3) gives

\begin{equation}
\begin{aligned}
f(q_+,q_-)&=f(-q_+,-q_-),\\
d(q_+,q_-)&=d(-q_+,-q_-),\\
f(q_+,q_-)&=f(-q_-,-q_+),\\
-d(q_+,q_-)&=d(-q_-,-q_+).
\end{aligned}
\end{equation}

Thus, the corresponding effective Hamiltonian retained to the lowest order of $f(\emph{\textbf{q}})$ and $d(\emph{\textbf{q}})$ can be simplified to
\begin{equation}
H_{eff}(\textbf{\emph{q}})=[a_1(q_+^2+q_-^2)+a_2q_+q_-]\sigma_z+\alpha(q_+^2-q_-^2)\sigma_++H.c..
\end{equation}
where $a_{1,2}$ is a real parameter. Thus, we conclude that the
doubly degenerate line at leading order is a QNL with an approximate
chiral symmetry.

\subsubsection{Point group $C_{2}$ with $S_4\mathcal{T}$}

Considering a double degenerate phonon band where the little group is $C_{2}$.
Considering the time reversal symmetry, one-dimensional irreducible representations can be combined into
a two-dimensional irreducible representation. So there is a
double degenerate band. $C_{2}$ contains one generator:
$C_{2z}$. The basis functions are $\{x,\mathcal{T}x\}$.

To construct the $k\cdot p$ effective Hamiltonian around a generic point, we
express the symmetry operators in the basis:

\begin{equation}
\begin{aligned}
&C_{2z}=\left(
\begin{array}
[c]{cc}%
-1 & 0\\
0 & -1
\end{array}
\right)  ,
C_{2z}^{-1}=\left(
\begin{array}
[c]{cc}%
-1 & 0\\
0 & -1
\end{array}
\right)  ,\\
\end{aligned}
\end{equation}

The constraint Eq. (S3) gives

\begin{equation}
\begin{aligned}
f(q_+,q_-)&=f(-q_+,-q_-),\\
d(q_+,q_-)&=d(-q_+,-q_-).
\end{aligned}
\end{equation}

Thus, the corresponding effective Hamiltonian retained to the lowest order of $f(\emph{\textbf{q}})$ and $d(\emph{\textbf{q}})$ can be simplified to
\begin{equation}
H_{eff}(\textbf{\emph{q}})=(a_1q_+^2+a_2q_-^2+a_3q_+q_-)\sigma_z+(\alpha_1q_+^2+\alpha_2q_-^2+\alpha_3q_+q_-)\sigma_++H.c..
\end{equation}
where $a_{1,2,3}$ is a real parameter,$\alpha_{1,2,3}$ is a complex parameter. Thus, we conclude that the
doubly degenerate line at leading order is a QNL with an approximate
chiral symmetry.

\subsubsection{Point group $C_{3}$ with $M_{z}\mathcal{T}$ symmetry}
Consider a double degenerate phonon band where the little group is $C_{3}$. We
know that point group $C_{3}$ has two one-dimensional
irreducible representations which are complex conjugate to each other. Considering the symmetry
of combined operator $M_{z}\mathcal{T}$, two one-dimensional irreducible representations can be combined into
a two-dimensional irreducible representation. So there is a
double degenerate band. The basis functions are $\{-(S_{x}+iS_{y}),(S_{x}%
-iS_{y})\}$

To construct the $k\cdot p$ effective Hamiltonian around a generic point, we
express the two symmetry operators $C_{3}$ and $M_{z}\mathcal{T}$ in the basis:

\begin{equation}
\begin{aligned}
&C_{3z}=\left(
\begin{array}
[c]{cc}%
e^{\frac{2i\pi}{3}} & 0\\
0 & e^{\frac{-2i\pi}{3}}%
\end{array}
\right)  ,
C_{4z}^{-1}=\left(
\begin{array}
[c]{cc}%
e^{\frac{-2i\pi}{3}} & 0\\
0 & e^{\frac{2i\pi}{3}}%
\end{array}
\right)  ,\\
&M_{z}\mathcal{T}=\sigma_{x}\mathcal{K}=\left(
\begin{array}
[c]{cc}%
0 & 1\\
1 & 0
\end{array}
\right) \mathcal{K},
(M_z\mathcal{T})^{-1}=\sigma_{x}\mathcal{K}=\left(
\begin{array}
[c]{cc}%
0 & 1\\
1 & 0
\end{array}
\right)  \mathcal{K},\\
\end{aligned}
\end{equation}

The constraint Eq. (S3) gives

\begin{equation}
\begin{aligned}
f(q_+,q_-)&=f(e^{i2\pi/3}q_+,e^{-i2\pi/3}q_-),\\
e^{i4\pi/3}d(q_+,q_-)&=d(e^{i2\pi/3}q_+,e^{-i2\pi/3}q_-),\\
-f^*(q_+,q_-)&=f(-q_+,-q_-),\\
d(q_+,q_-)&=d(-q_+,-q_-).
\end{aligned}
\end{equation}

Thus, the corresponding effective Hamiltonian retained to the lowest order of $f(\emph{\textbf{q}})$ and $d(\emph{\textbf{q}})$ can be simplified to
\begin{equation}
H_{eff}(\textbf{\emph{q}})=a(q_+^3+q_-^3)\sigma_z+\alpha q_+^2\sigma_++H.c..
\end{equation}

Thus, we conclude that the doubly degenerate line at leading order is a QNL with an approximate
chiral symmetry.

\subsubsection{Point group $C_{3v}$with $M_{z}\mathcal{T}$ symmetry}

Considering a double degenerate phonon band where the little group is $C_{3v}$.
We know that point group $C_{3v}$ has a two-dimensional irreducible
representation which corresponds to a doubly degenerate band.
$C_{3v}$ contains two generators: $C_{3z}$, $M_{y}$. The basis
functions are $\{(S_{x}-iS_{y}),-(S_{x}+iS_{y})\}$

To construct the $k\cdot p$ effective Hamiltonian around a generic point, we
express the three symmetry operators $C_{3z}$, $M_{y}$, and $M_{z}T$  in the basis:

 \begin{equation}
 \begin{aligned}
 &C_{3z}=\left(
\begin{array}
[c]{cc}%
e^{\frac{2i\pi}{3}} & 0\\
0 & e^{\frac{-2i\pi}{3}}%
\end{array}
\right)  ,
C_{3z}^{-1}=\left(
\begin{array}
[c]{cc}%
e^{\frac{-2i\pi}{3}} & 0\\
0 & e^{\frac{2i\pi}{3}}%
\end{array}
\right)  ,\\
&M_{x}=\left(
\begin{array}
[c]{cc}%
0 & 1\\
1 & 0
\end{array}
\right)  ,M_{x}^{-1}=\left(
\begin{array}
[c]{cc}%
0 & 1\\
1 & 0
\end{array}
\right)  ,\\
&M_{z}\mathcal{T}=\sigma_{x}\mathcal{K}=\left(
\begin{array}
[c]{cc}%
0 & 1\\
1 & 0
\end{array}
\right)  \mathcal{K},
(M_{z}\mathcal{T})^{-1}=\sigma_{x}\mathcal{K}=\left(
\begin{array}
[c]{cc}%
0 & 1\\
1 & 0
\end{array}
\right)  \mathcal{K},\\
\end{aligned}
\end{equation}

The constraint Eq. (S3) gives

\begin{equation}
\begin{aligned}
f(q_+,q_-)&=f(e^{i2\pi/3}q_+,e^{-i2\pi/3}q_-),\\
e^{i4\pi/3}d(q_+,q_-)&=d(e^{i2\pi/3}q_+,e^{-i2\pi/3}q_-),\\
-f(q_+,q_-)&=f(-q_-,-q_+),\\
d^*(q_+,q_-)&=d(-q_-,-q_+)\\
-f^*(q_+,q_-)&=f(-q_+,-q_-),\\
d(q_+,q_-)&=d(-q_+,-q_-).
\end{aligned}
\end{equation}

Thus, the corresponding effective Hamiltonian retained to the lowest order of $f(\emph{\textbf{q}})$ and $d(\emph{\textbf{q}})$ can be simplified to
\begin{equation}
H_{eff}(\textbf{\emph{q}})=a_1(q_+^3+q_-^3)\sigma_z+a_2q_+^2\sigma_++H.c..
\end{equation}
where $a_{1,2}$ is a real. Thus, we conclude that the doubly degenerate line at leading order is a QNL with an approximate
chiral symmetry.

\subsubsection{ Point Group $C_{4}$ with $\mathcal{PT}$ symmetry}

Considering a double degenerate phonon band where the little group is $C_{4}$. We
know that point group $C_{4}$ has two one-dimensional
irreducible representations which are complex conjugate to each other. Considering the symmetry
of combined operator $\mathcal{PT}$, two one-dimensional irreducible representations can be combined into
a two-dimensional irreducible representation. So there is a
double degenerate band. $C_{4}$ contains one generator: $C_{4z}$.
The basis functions are $\{-i(x+iy),i(x-iy)\}$ or $\{-(S_{x}+iS_{y}),(S_{x}-iS_{y})\}$

To construct the $k\cdot p$ effective Hamiltonian around a generic point, we
express the symmetry operators $C_{4z}$, $\mathcal{PT}$ in the basis:

\begin{equation}
\begin{aligned}
&C_{4z}=\left(
\begin{array}
[c]{cc}%
i & 0\\
0 & -i
\end{array}
\right)  ,
C_{4z}^{-1}=\left(
\begin{array}
[c]{cc}%
-i & 0\\
0 & i
\end{array}
\right)  ,\\
&\mathcal{PT}=\left(
\begin{array}
[c]{cc}%
0 & 1\\
1 & 0
\end{array}
\right)  \mathcal{K},
 (\mathcal{PT})^{-1}=\left(
\begin{array}
[c]{cc}%
0 & 1\\
1 & 0
\end{array}
\right)  \mathcal{K},\\
\end{aligned}
\end{equation}

The constraint Eq. (S3) gives

\begin{equation}
\begin{aligned}
f(q_+,q_-)&=f(iq_+,iq_-),\\
-d(q_+,q_-)&=d(iq_+,iq_-),\\
-f^*(q_+,q_-)&=f(q_+,q_-),\\
d(q_+,q_-)&=d(q_+,q_-).
\end{aligned}
\end{equation}

Thus, the corresponding effective Hamiltonian retained to the lowest order of $f(\emph{\textbf{q}})$ and $d(\emph{\textbf{q}})$ can be simplified to

\begin{equation}
H_{eff}(\textbf{\emph{q}})=a(q_+^4-q_-^4)\sigma_z+(\alpha_1q_+^2+\alpha_2q_-^2)\sigma_++H.c..
\end{equation}

where $\alpha_{1,2}$ is a complex. Thus, we conclude that the doubly degenerate line at leading order is a QNL with an approximate
chiral symmetry.

\subsubsection{Point Group $C_{4v}$}

Considering a double degenerate phonon band where the little group is $C_{4v}$.
We know that point group $C_{4v}$ has a two-dimensional irreducible
representation which corresponding to a double degenerate band. $C_{4v}$ contains two generators: $C_{4z},M_{x}$. The basis
functions are $\{S_{x},S_{y}\}$.

To construct the $k\cdot p$ effective Hamiltonian around a generic point, we
express the symmetry operators in the basis:

\begin{equation}
\begin{aligned}
&C_{4z}=\left(
\begin{array}
[c]{cc}%
0 & -1\\
1 & 0
\end{array}
\right)  ,
C_{4z}^{-1}=\left(
\begin{array}
[c]{cc}%
0 & 1\\
-1 & 0
\end{array}
\right)  ,\\
&M_{x}=\left(
\begin{array}
[c]{cc}%
-1 & 0\\
0 & 1
\end{array}
\right)  ,
M_{x}^{-1}=\left(
\begin{array}
[c]{cc}%
-1 & 0\\
0 & 1
\end{array}
\right)  ,\\
\end{aligned}
\end{equation}

The constraint Eq. (S3) gives

\begin{equation}
\begin{aligned}
-f(q_+,q_-)&=f(iq_+,iq_-),\\
-d^*(q_+,q_-)&=d(iq_+,iq_-),\\
f(q_+,q_-)&=f(-q_-,-q_+),\\
-d(q_+,q_-)&=d(-q_-,-q_+).
\end{aligned}
\end{equation}

Thus, the corresponding effective Hamiltonian retained to the lowest order of $f(\emph{\textbf{q}})$ and $d(\emph{\textbf{q}})$ can be simplified to

\begin{equation}
H_{eff}(\textbf{\emph{q}})=[a_1(q_+^2+q_-^2)+a_2q_+q_-]\sigma_z+ia_3(q_+^2-q_-^2)\sigma_++H.c..
\end{equation}

where $a_{1,2,3}$ is a real parameter. Thus, we conclude that the doubly degenerate line at leading order is a QNL with an approximate
chiral symmetry.

\subsubsection{Point Group $C_{6}$ with $\mathcal{PT}$}

Considering a double degenerate phonon band where the little group is $C_{6}$. We
know that point group $C_{6}$ has two one-dimensional
 irreducible representations which are complex conjugate to each other. Considering the symmetry
of combined operator $\mathcal{PT}$, two one-dimensional irreducible representations can be combined into
a two-dimensional irreducible representation. So there is a
double degenerate band. $C_{6}$ contains one generator: $C_{6z}$.
The basis functions are $\{(x-iy)^{2},(x+iy)^{2}\}$ or $\{-(S_{x}+iS_{y}),(S_{x}-iS_{y})\}$

To construct the $k\cdot p$ effective Hamiltonian around a generic
point, we express the two symmetry operators  $C_{6z}$, $\mathcal{PT}$ in the basis:

\begin{equation}
\begin{aligned}
&C_{6z}=\left(
\begin{array}
[c]{cc}%
e^{\frac{i\pi}{3}} & 0\\
0 & e^{\frac{-i\pi}{3}}%
\end{array}
\right)  ,C_{6z}^{-1}=\left(
\begin{array}
[c]{cc}%
e^{\frac{-i\pi}{3}} & 0\\
0 & e^{\frac{i\pi}{3}}%
\end{array}
\right)  ,\\
&\mathcal{PT}=\sigma_{x}\mathcal{K}=\left(
\begin{array}
[c]{cc}%
0 & 1\\
1 & 0
\end{array}
\right)  \mathcal{K},
(\mathcal{PT})^{-1}=\sigma_{x}\mathcal{K}=\left(
\begin{array}
[c]{cc}%
0 & 1\\
1 & 0
\end{array}
\right)  \mathcal{K},\\
\end{aligned}
\end{equation}

The constraint Eq. (S3) gives

\begin{equation}
\begin{aligned}
f(q_+,q_-)&=f(e^{i\pi/3}q_+,e^{-i\pi/3}q_-),\\
e^{i2\pi/3}d(q_+,q_-)&=d(e^{i\pi/3}q_+,e^{-i\pi/3}q_-),\\
-f^*(q_+,q_-)&=f(q_+,q_-),\\
d^(q_+,q_-)&=d(q_+,q_-).
\end{aligned}
\end{equation}

Thus, the corresponding effective Hamiltonian retained to the lowest order of $f(\emph{\textbf{q}})$ and $d(\emph{\textbf{q}})$ can be simplified to

\begin{equation}
H_{eff}(\textbf{\emph{q}})=a(q_+^6-q_-^6)\sigma_z+\alpha q_+^2\sigma_++H.c..
\end{equation}

Thus, we conclude that the doubly degenerate line at leading order is a QNL with an approximate
chiral symmetry.

\subsubsection{Point group $C_{6v}$}

Considering a double degenerate phonon band where the little group is $C_{6v}$.
We know that point group $C_{6v}$ has a two-dimensional irreducible
representation which corresponding a double degenerate band.
$C_{6v}$ contains two generators: $C_{6z},M_{x}$. And the basis functions are
$\{(S_{x}-iS_{y}),-(S_{x}+iS_{y})\}$

To construct the $k\cdot p$ effective Hamiltonian around a generic point, we
express the symmetry operators in the basis:

\begin{equation}
\begin{aligned}
&C_{6z}=\left(
\begin{array}
[c]{cc}%
e^{\frac{i\pi}{3}} & 0\\
0 & e^{\frac{-i\pi}{3}}%
\end{array}
\right)  ,
C_{6z}^{-1}=\left(
\begin{array}
[c]{cc}%
e^{\frac{-i\pi}{3}} & 0\\
0 & e^{\frac{i\pi}{3}}%
\end{array}
\right)  ,\\
&M_{x}=\left(
\begin{array}
[c]{cc}%
0 & 1\\
1 & 0
\end{array}
\right)  ,
M_{x}^{-1}=\left(
\begin{array}
[c]{cc}%
0 & 1\\
1 & 0
\end{array}
\right)  ,\\
\end{aligned}
\end{equation}

The constraint Eq. (S3) gives

\begin{equation}
\begin{aligned}
f(q_+,q_-)&=f(e^{i\pi/3}q_+,e^{-i\pi/3}q_-),\\
e^{i2\pi/3}d(q_+,q_-)&=d(e^{i\pi/3}q_+,e^{-i\pi/3}q_-),\\
-f(q_+,q_-)&=f(-q_-,-q_+),\\
d^*(q_+,q_-)&=d(-q_-,-q_+).
\end{aligned}
\end{equation}

Thus, the corresponding effective Hamiltonian retained to the lowest order of $f(\emph{\textbf{q}})$ and $d(\emph{\textbf{q}})$ can be simplified to

\begin{equation}
H_{eff}(\textbf{\emph{q}})=a_1(q_+^6-q_-^6)\sigma_z+a_2q_+^2\sigma_++H.c..
\end{equation}
where $a_{1,2}$ is a real parameter. Thus, we conclude that the doubly degenerate line at leading order is a QNL with an approximate
chiral symmetry.

\section{Nodal lines Table}

\linespread{1}
\begin{table*}[!htbp]
\footnotesize  
\centering
\setlength{\tabcolsep}{2mm}
\renewcommand{\thetable}{S\arabic{table}}
\caption{List of all possible NLs stabilized by point group symmetry. Location represents a high symmetry line with double degenerate nodal lines, and little group represents the corresponding wave vector group.}
\begin{tabular}{ccccccc}
\hline 
\ space group& location &little group &basis function &berry phase &generator  \\
\hline
\ 147	&(0,0,w)      &$C_3$	   &$\{-i(x+iy),i(x-iy)\}$	          &$\pi$	&$C_{3z},\mathcal{PT}$\\
\ 147   &(1/3,1/3,w)  &$C_3$       &$\{-i(x+iy),i(x-iy)\}$	          &$\pi$	&$C_{3z},\mathcal{PT}$\\
\ 148	&(w,w,w)	  &$C_3$       &$\{-i(x+iy),i(x-iy)\}$          &$\pi$    &$C_{3z},\mathcal{PT}$\\
\ 156	&(0,0,w)	  &$C_{3v}$	   &$\{(S_x-iS_y),-(S_x+iS_y)\}$  &$\pi$    &$C_{3z},M_x$\\
\ 157	&(0,0,w)	  &$C_{3v}$	   &$\{(S_x-iS_y),-(S_x+iS_y)\}$  &$\pi$    &$C_{3z},M_x$\\
\ 157	&(1/3,1/3,w)  &$C_{3v}$    &$\{(S_x-iS_y),-(S_x+iS_y)\}$  &$\pi$    &$C_{3z},M_x$\\
\ 157	&(-1/3,-1/3,w)&$C_{3v}$    &$\{(S_x-iS_y),-(S_x+iS_y)\}$  &-$\pi$   &$C_{3z},M_x$\\
\ 160	&(w,w,w)	  &$C_{3v}$	   &$\{(S_x-iS_y),-(S_x+iS_y)\}$  &$\pi$	&$C_{3z},M_x$\\
\ 162	&(0,0,w)	  &$C_{3v}$	   &$\{(S_x-iS_y),-(S_x+iS_y)\}$  &$\pi$	&$C_{3z},M_x,\mathcal{PT}$\\
\ 162	&(1/3,1/3,w)  &$C_{3v}$    &$\{(S_x-iS_y),-(S_x+iS_y)\}$  &$\pi$	&$C_{3z},M_x,\mathcal{PT}$\\
\ 164	&(0,0,w)	  &$C_{3v}$	   &$\{(S_x-iS_y),-(S_x+iS_y)\}$  &$\pi$	&$C_{3z},M_x,\mathcal{PT}$\\
\ 164	&(1/3,1/3,w)  &$C_3$	   &$\{-i(x+iy),i(x-iy)\}$	          &$\pi$    &$C_{3z},\mathcal{PT}$\\
\ 166	&(0,0,w)	  &$C_{3v}$	   &$\{(S_x-iS_y),-(S_x+iS_y)\}$  &$\pi$	&$C_{3z},M_x,\mathcal{PT}$\\
\ 175	&(1/3,1/3,w)  &$C_3$       &$\{-i(x+iy),i(x-iy)\}$             &$\pi$    &$C_{3z},\mathcal{PT}$\\
\ 183	&(1/3,1/3,w)  &$C_{3v}$	   &$\{(S_x-iS_y),-(S_x+iS_y)\}$  &$\pi$	&$C_{3z},M_x$\\
\ 189	&(1/3,1/3,w)  &$C_{3v}$	   &$\{(S_x-iS_y),-(S_x+iS_y)\}$  &$\pi$    &$C_{3z},M_x$\\
\ 189	&(-1/3,-1/3,w)&$C_{3v}$    &$\{(S_x-iS_y),-(S_x+iS_y)\}$  &-$\pi$   &$C_{3z},M_x$\\
\ 191	&(1/3,1/3,w)  &$C_{3v}$	   &$\{(S_x-iS_y),-(S_x+iS_y)\}$  &$\pi$    &$C_{3z},M_x,PT$\\
\ 200	&(w,w,w)	  &$C_3$	   &$\{-i(x+iy),i(x-iy)\}$	          &$\pi$	&$C_{3z},\mathcal{PT}$\\
\ 202   &(w,w,w)	  &$C_3$	   &$\{-i(x+iy),i(x-iy)\}$	          &$\pi$    &$C_{3z},\mathcal{PT}$\\
\ 204   &(w,w,w)	  &$C_3$	   &$\{-i(x+iy),i(x-iy)\}$	          &$\pi$    &$C_{3z},\mathcal{PT}$\\
\ 215	&(w,w,w)	  &$C_{3v}$	   &$\{(S_x-iS_y),-(S_x+iS_y)\}$  &$\pi$	&$C_{3z},M_x$\\
\ 216	&(w,w,w)	  &$C_{3v}$    &$\{(S_x-iS_y),-(S_x+iS_y)\}$  &$\pi$	&$C_{3z},M_x$\\
\ 217   &(w,w,w)	  &$C_{3v}$	   &$\{(S_x-iS_y),-(S_x+iS_y)\}$  &$\pi$	&$C_{3z},M_x$\\
\ 221	&(w,w,w)	  &$C_{3v}$	   &$\{(S_x-iS_y),-(S_x+iS_y)\}$  &$\pi$	&$C_{3z},M_x,\mathcal{PT}$\\
\ 225	&(w,w,w)	  &$C_{3v}$	   &$\{(S_x-iS_y),-(S_x+iS_y)\}$  &$\pi$	&$C_{3z},M_x,\mathcal{PT}$\\
\ 229	&(w,w,w)	  &$C_{3v}$	   &$\{(S_x-iS_y),-(S_x+iS_y)\}$  &$\pi$	&$C_{3z},M_x,\mathcal{PT}$\\\\
\ 81	&(0,0,w)	  &$C_2$	   &$\{x,\mathcal{T}x\}$	          &2$\pi$   &$C_2,\mathcal{S}_4\mathcal{T}$\\
\ 83	&(0,0,w)	  &$C_4$	   &$\{-i(x+iy),i(x-iy)\}$	          &2$\pi$   &$C_{4z},\mathcal{PT}$\\
\ 83	&(1/2,1/2,w)  &$C_4$       &$\{-i(x+iy),i(x-iy)\}$         &2$\pi$   &$C_{4z},\mathcal{PT}$\\
\ 87	&(w,w,-w)	  &$C_4$	   &$\{-i(x+iy),i(x-iy)\}$             &2$\pi$   &$C_{4z},\mathcal{PT}$\\
\ 99	&(0,0,w)	  &$C_{4v}$    &$\{S_x,S_y\}$                 &2$\pi$   &$C_{4z},M_x$\\
\ 99	&(1/2,1/2,w)  &	$C_{4v}$   &$\{S_x,S_y\}$                 &2$\pi$	&$C_{4z},M_x$\\
\ 107	&(w,w,-w)	  &$C_{4v}$    &$\{S_x,S_y\}$	              &2$\pi$   &$C_{4z},M_x$\\
\ 111	&(0,0,w)	  &$C_{2v}$    &$\{x,y\}$	                          &2$\pi$	&$C_{2z},M_x,\mathcal{S}_4\mathcal{T}$\\
\ 111	&(1/2,1/2,w)  &	$C_{2v}$   &$\{x,y\}$	                         &2$\pi$	&$C_{2z},M_x,\mathcal{S}_4\mathcal{T}$\\
\ 115	&(0,0,w)	  &$C_{2v}$	   &$\{x,y\}$	                          &	2$\pi$  &$C_{2z},M_x,\mathcal{S}_4\mathcal{T}$\\
\ 115	&(1/2,1/2,w)  &$C_{2v}$	   &$\{x,y\}$	                          &2$\pi$   &$C_{2z},M_x,\mathcal{S}_4\mathcal{T}$\\
\ 119	&(w,w,-w)	  &$C_{2v}$    &$\{x,y\}$		                          &2$\pi$	&$C_{2z},M_x,\mathcal{S}_4\mathcal{T}$\\
\ 121	&(w,w,-w)	  &$C_{2v}$	   &$\{x,y\}$	                         &2$\pi$	&$C_{2z},M_x,\mathcal{S}_4\mathcal{T}$\\
\ 123	&(0,0,w)	  &$C_{4v}$	   &$\{S_x,S_y\}$	              &2$\pi$   &$C_{4z},M_x,\mathcal{PT}$\\
\ 123	&(1/2,1/2,w)  &	$C_{4v}$   &$\{S_x,S_y\}$	              &2$\pi$	&$C_{4z},M_x,\mathcal{PT}$\\
\ 139	&(w,w,-w)	  &$C_{4v}$    &$\{S_x,S_y\}$	              &2$\pi$   &$C_{4z},M_x,\mathcal{PT}$\\
\ 174	&(0,0,w)	  &$C_3$       &$\{-i(x+iy),i(x-iy)\}$          &2$\pi$   &$C_3,M_zT$\\
\  175	&(0,0,w)	  &$C_6$	   &$\{(x-iy)^2,(x+iy)^2\}$	              &2$\pi$   &$C_6,M_z\mathcal{T},\mathcal{PT}$\\
\ 183	&(0,0,w)	  &$C_{6v}$    &$\{(S_x-iS_y),-(S_x+iS_y)\}$  &	2$\pi$  &$C_{6z},M_x$\\
\ 187	&(0,0,w)	  &$C_{3v}$	   &$\{(S_x-iS_y),-(S_x+iS_y)\}$   &2$\pi$  &$C_{3z},M_x,M_z\mathcal{T}$\\
\ 189	&(0,0,w)	  &$C_{3v}$    &$\{(S_x-iS_y),-(S_x+iS_y)\}$   &2$\pi$  &$C_{3z},M_x,M_z\mathcal{T}$\\
\ 191	&(0,0,w)	  &$C_{6v}$    &$\{(S_x-iS_y),-(S_x+iS_y)\}$   &2$\pi$  &$C_{6z},M_x,\mathcal{PT}$\\
\ 215	&(1/2,1/2,w)  &$C_{2v}$    &$\{x,y\}$	                           &2$\pi$	&$C_{2z},M_x,\mathcal{S}_4\mathcal{T}$\\
\ 215	&(0,w,0)	  &$C_{2v}$    &$\{x,y\}$	                          &2$\pi$	&$C_{2z},M_x,\mathcal{S}_4\mathcal{T}$\\
\ 216	&(w,w,0)	  &$C_{2v}$    &$\{x,y\}$	                         &2$\pi$	&$C_{2z},M_x,\mathcal{S}_4\mathcal{T}$\\
\ 217	&(w,-w,w)	  &$C_{2v}$    &$\{x,y\}$	                           &2$\pi$  &$C_{2z},M_x,\mathcal{S}_4\mathcal{T}$\\
\ 221	&(1/2,1/2,w)  &$C_{4v}$    &$\{S_x,S_y\}$                  &2$\pi$  &$C_{4z},M_x,\mathcal{PT}$\\
\ 221	&(0,w,0)      &$C_{4v}$    &$\{S_x,S_y\}$                  &2$\pi$	&$C_{4z},M_x,\mathcal{PT}$\\
\ 225	&(w,w,0)      &$C_{4v}$    &$\{S_x,S_y\}$                  &2$\pi$	&$C_{4z},M_x,\mathcal{PT}$\\
\ 229	&(w,-w,w)     &$C_{4v}$    &$\{S_x,S_y\}$                  &2$\pi$  &$C_{4z},M_x,\mathcal{PT}$\\
\hline 
\end{tabular}
\end{table*}


\begin{thebibliography}{44}
\expandafter\ifx\csname natexlab\endcsname\relax\def\natexlab#1{#1}\fi
\expandafter\ifx\csname bibnamefont\endcsname\relax
  \def\bibnamefont#1{#1}\fi
\expandafter\ifx\csname bibfnamefont\endcsname\relax
  \def\bibfnamefont#1{#1}\fi
\expandafter\ifx\csname citenamefont\endcsname\relax
  \def\citenamefont#1{#1}\fi
\expandafter\ifx\csname url\endcsname\relax
  \def\url#1{\texttt{#1}}\fi
\expandafter\ifx\csname urlprefix\endcsname\relax\def\urlprefix{URL }\fi
\providecommand{\bibinfo}[2]{#2}
\providecommand{\eprint}[2][]{\url{#2}}

\bibitem[{\citenamefont{Hasan and Kane}(2010)}]{RevModPhys.82.3045}
\bibinfo{author}{\bibfnamefont{M.~Z.} \bibnamefont{Hasan}} \bibnamefont{and}
  \bibinfo{author}{\bibfnamefont{C.~L.} \bibnamefont{Kane}},
  \bibinfo{journal}{Rev. Mod. Phys.} \textbf{\bibinfo{volume}{82}},
  \bibinfo{pages}{3045} (\bibinfo{year}{2010}).

\bibitem[{\citenamefont{Qi and Zhang}(2011)}]{RevModPhys.83.1057}
\bibinfo{author}{\bibfnamefont{X.~L.} \bibnamefont{Qi}} \bibnamefont{and}
  \bibinfo{author}{\bibfnamefont{S.~C.} \bibnamefont{Zhang}},
  \bibinfo{journal}{Rev. Mod. Phys.} \textbf{\bibinfo{volume}{83}},
  \bibinfo{pages}{1057} (\bibinfo{year}{2011}).

\bibitem[{\citenamefont{Armitage et~al.}(2018)\citenamefont{Armitage, Mele, and
  Vishwanath}}]{RevModPhys.90.015001}
\bibinfo{author}{\bibfnamefont{N.~P.} \bibnamefont{Armitage}},
  \bibinfo{author}{\bibfnamefont{E.~J.} \bibnamefont{Mele}}, \bibnamefont{and}
  \bibinfo{author}{\bibfnamefont{A.}~\bibnamefont{Vishwanath}},
  \bibinfo{journal}{Rev. Mod. Phys.} \textbf{\bibinfo{volume}{90}},
  \bibinfo{pages}{015001} (\bibinfo{year}{2018}).

\bibitem[{\citenamefont{Wang et~al.}(2012)\citenamefont{Wang, Sun, Chen,
  Franchini, Xu, Weng, Dai, and Fang}}]{PhysRevB.85.195320}
\bibinfo{author}{\bibfnamefont{Z.}~\bibnamefont{Wang}},
  \bibinfo{author}{\bibfnamefont{Y.}~\bibnamefont{Sun}},
  \bibinfo{author}{\bibfnamefont{X.~Q.} \bibnamefont{Chen}},
  \bibinfo{author}{\bibfnamefont{C.}~\bibnamefont{Franchini}},
  \bibinfo{author}{\bibfnamefont{G.}~\bibnamefont{Xu}},
  \bibinfo{author}{\bibfnamefont{H.}~\bibnamefont{Weng}},
  \bibinfo{author}{\bibfnamefont{X.}~\bibnamefont{Dai}}, \bibnamefont{and}
  \bibinfo{author}{\bibfnamefont{Z.}~\bibnamefont{Fang}},
  \bibinfo{journal}{Phys. Rev. B} \textbf{\bibinfo{volume}{85}},
  \bibinfo{pages}{195320} (\bibinfo{year}{2012}).

\bibitem[{\citenamefont{Wan et~al.}(2011)\citenamefont{Wan, Turner, Vishwanath,
  and Savrasov}}]{PhysRevB.83.205101}
\bibinfo{author}{\bibfnamefont{X.}~\bibnamefont{Wan}},
  \bibinfo{author}{\bibfnamefont{A.~M.} \bibnamefont{Turner}},
  \bibinfo{author}{\bibfnamefont{A.}~\bibnamefont{Vishwanath}},
  \bibnamefont{and} \bibinfo{author}{\bibfnamefont{S.~Y.}
  \bibnamefont{Savrasov}}, \bibinfo{journal}{Phys. Rev. B}
  \textbf{\bibinfo{volume}{83}}, \bibinfo{pages}{205101}
  (\bibinfo{year}{2011}).

\bibitem[{\citenamefont{Weng et~al.}(2015)\citenamefont{Weng, Fang, Fang,
  Bernevig, and Dai}}]{weng2015weyl}
\bibinfo{author}{\bibfnamefont{H.}~\bibnamefont{Weng}},
  \bibinfo{author}{\bibfnamefont{C.}~\bibnamefont{Fang}},
  \bibinfo{author}{\bibfnamefont{Z.}~\bibnamefont{Fang}},
  \bibinfo{author}{\bibfnamefont{B.~A.} \bibnamefont{Bernevig}},
  \bibnamefont{and} \bibinfo{author}{\bibfnamefont{X.}~\bibnamefont{Dai}},
  \bibinfo{journal}{Physical Review X} \textbf{\bibinfo{volume}{5}},
  \bibinfo{pages}{011029} (\bibinfo{year}{2015}).

\bibitem[{\citenamefont{Wang et~al.}(2019)\citenamefont{Wang, Ruan, and
  Zhang}}]{PhysRevB.99.075130}
\bibinfo{author}{\bibfnamefont{H.}~\bibnamefont{Wang}},
  \bibinfo{author}{\bibfnamefont{J.}~\bibnamefont{Ruan}}, \bibnamefont{and}
  \bibinfo{author}{\bibfnamefont{H.}~\bibnamefont{Zhang}},
  \bibinfo{journal}{Phys. Rev. B} \textbf{\bibinfo{volume}{99}},
  \bibinfo{pages}{075130} (\bibinfo{year}{2019}).

\bibitem[{\citenamefont{Yu et~al.}(2019)\citenamefont{Yu, Wu, Sheng, Zhao, and
  Yang}}]{PhysRevB.99.121106}
\bibinfo{author}{\bibfnamefont{Z.M.} \bibnamefont{Yu}},
  \bibinfo{author}{\bibfnamefont{W.}~\bibnamefont{Wu}},
  \bibinfo{author}{\bibfnamefont{X.~L.} \bibnamefont{Sheng}},
  \bibinfo{author}{\bibfnamefont{Y.~X.} \bibnamefont{Zhao}}, \bibnamefont{and}
  \bibinfo{author}{\bibfnamefont{S.~A.} \bibnamefont{Yang}},
  \bibinfo{journal}{Phys. Rev. B} \textbf{\bibinfo{volume}{99}},
  \bibinfo{pages}{121106} (\bibinfo{year}{2019}).

\bibitem[{\citenamefont{Bian et~al.}(2016)\citenamefont{Bian, Chang, Zheng,
  Velury, Xu, Neupert, Chiu, Huang, Sanchez, Belopolski
  et~al.}}]{PhysRevB.93.121113}
\bibinfo{author}{\bibfnamefont{G.}~\bibnamefont{Bian}},
  \bibinfo{author}{\bibfnamefont{T.R.} \bibnamefont{Chang}},
  \bibinfo{author}{\bibfnamefont{H.}~\bibnamefont{Zheng}},
  \bibinfo{author}{\bibfnamefont{S.}~\bibnamefont{Velury}},
  \bibinfo{author}{\bibfnamefont{S.~Y.} \bibnamefont{Xu}},
  \bibinfo{author}{\bibfnamefont{T.}~\bibnamefont{Neupert}},
  \bibinfo{author}{\bibfnamefont{C.~K.} \bibnamefont{Chiu}},
  \bibinfo{author}{\bibfnamefont{S.~M.} \bibnamefont{Huang}},
  \bibinfo{author}{\bibfnamefont{D.~S.} \bibnamefont{Sanchez}},
  \bibinfo{author}{\bibfnamefont{I.}~\bibnamefont{Belopolski}},
  \bibnamefont{et~al.}, \bibinfo{journal}{Phys. Rev. B}
  \textbf{\bibinfo{volume}{93}}, \bibinfo{pages}{121113}
  (\bibinfo{year}{2016}).

\bibitem[{\citenamefont{Fang et~al.}(2015)\citenamefont{Fang, Chen, Kee, and
  Fu}}]{PhysRevB.92.081201}
\bibinfo{author}{\bibfnamefont{C.}~\bibnamefont{Fang}},
  \bibinfo{author}{\bibfnamefont{Y.}~\bibnamefont{Chen}},
  \bibinfo{author}{\bibfnamefont{H.~Y.} \bibnamefont{Kee}}, \bibnamefont{and}
  \bibinfo{author}{\bibfnamefont{L.}~\bibnamefont{Fu}}, \bibinfo{journal}{Phys.
  Rev. B} \textbf{\bibinfo{volume}{92}}, \bibinfo{pages}{081201}
  (\bibinfo{year}{2015}).

\bibitem[{\citenamefont{Hirayama et~al.}(2017)\citenamefont{Hirayama, Okugawa,
  Miyake, and Murakami}}]{hirayama2017topological}
\bibinfo{author}{\bibfnamefont{M.}~\bibnamefont{Hirayama}},
  \bibinfo{author}{\bibfnamefont{R.}~\bibnamefont{Okugawa}},
  \bibinfo{author}{\bibfnamefont{T.}~\bibnamefont{Miyake}}, \bibnamefont{and}
  \bibinfo{author}{\bibfnamefont{S.}~\bibnamefont{Murakami}},
  \bibinfo{journal}{Nature communications} \textbf{\bibinfo{volume}{8}},
  \bibinfo{pages}{1} (\bibinfo{year}{2017}).

\bibitem[{\citenamefont{Yan et~al.}(2017)\citenamefont{Yan, Bi, Shen, Lu,
  Zhang, and Wang}}]{yan2017nodal}
\bibinfo{author}{\bibfnamefont{Z.}~\bibnamefont{Yan}},
  \bibinfo{author}{\bibfnamefont{R.}~\bibnamefont{Bi}},
  \bibinfo{author}{\bibfnamefont{H.}~\bibnamefont{Shen}},
  \bibinfo{author}{\bibfnamefont{L.}~\bibnamefont{Lu}},
  \bibinfo{author}{\bibfnamefont{S.~C.} \bibnamefont{Zhang}}, \bibnamefont{and}
  \bibinfo{author}{\bibfnamefont{Z.}~\bibnamefont{Wang}},
  \bibinfo{journal}{Physical Review B} \textbf{\bibinfo{volume}{96}},
  \bibinfo{pages}{041103} (\bibinfo{year}{2017}).

\bibitem[{\citenamefont{Huh et~al.}(2016)\citenamefont{Huh, Moon, and
  Kim}}]{PhysRevB.93.035138}
\bibinfo{author}{\bibfnamefont{Y.}~\bibnamefont{Huh}},
  \bibinfo{author}{\bibfnamefont{E.~G.} \bibnamefont{Moon}}, \bibnamefont{and}
  \bibinfo{author}{\bibfnamefont{Y.~B.} \bibnamefont{Kim}},
  \bibinfo{journal}{Phys. Rev. B} \textbf{\bibinfo{volume}{93}},
  \bibinfo{pages}{035138} (\bibinfo{year}{2016}).

\bibitem[{\citenamefont{Deng et~al.}(2019)\citenamefont{Deng, Lu, Li, Huang,
  Yan, Ma, and Liu}}]{deng2019nodal}
\bibinfo{author}{\bibfnamefont{W.}~\bibnamefont{Deng}},
  \bibinfo{author}{\bibfnamefont{J.}~\bibnamefont{Lu}},
  \bibinfo{author}{\bibfnamefont{F.}~\bibnamefont{Li}},
  \bibinfo{author}{\bibfnamefont{X.}~\bibnamefont{Huang}},
  \bibinfo{author}{\bibfnamefont{M.}~\bibnamefont{Yan}},
  \bibinfo{author}{\bibfnamefont{J.}~\bibnamefont{Ma}}, \bibnamefont{and}
  \bibinfo{author}{\bibfnamefont{Z.}~\bibnamefont{Liu}},
  \bibinfo{journal}{Nature communications} \textbf{\bibinfo{volume}{10}},
  \bibinfo{pages}{1} (\bibinfo{year}{2019}).

\bibitem[{\citenamefont{Zhong et~al.}(2016)\citenamefont{Zhong, Chen, Xie,
  Yang, Cohen, and Zhang}}]{C6NR00882H}
\bibinfo{author}{\bibfnamefont{C.}~\bibnamefont{Zhong}},
  \bibinfo{author}{\bibfnamefont{Y.}~\bibnamefont{Chen}},
  \bibinfo{author}{\bibfnamefont{Y.}~\bibnamefont{Xie}},
  \bibinfo{author}{\bibfnamefont{S.~A.} \bibnamefont{Yang}},
  \bibinfo{author}{\bibfnamefont{M.~L.} \bibnamefont{Cohen}}, \bibnamefont{and}
  \bibinfo{author}{\bibfnamefont{S.~B.} \bibnamefont{Zhang}},
  \bibinfo{journal}{Nanoscale} \textbf{\bibinfo{volume}{8}},
  \bibinfo{pages}{7232} (\bibinfo{year}{2016}).

\bibitem[{\citenamefont{Wu et~al.}(2018)\citenamefont{Wu, Liu, Li, Zhong, Yu,
  Sheng, Zhao, and Yang}}]{PhysRevB.97.115125}
\bibinfo{author}{\bibfnamefont{W.}~\bibnamefont{Wu}},
  \bibinfo{author}{\bibfnamefont{Y.}~\bibnamefont{Liu}},
  \bibinfo{author}{\bibfnamefont{S.}~\bibnamefont{Li}},
  \bibinfo{author}{\bibfnamefont{C.}~\bibnamefont{Zhong}},
  \bibinfo{author}{\bibfnamefont{Z.~M.} \bibnamefont{Yu}},
  \bibinfo{author}{\bibfnamefont{X.~L.} \bibnamefont{Sheng}},
  \bibinfo{author}{\bibfnamefont{Y.~X.} \bibnamefont{Zhao}}, \bibnamefont{and}
  \bibinfo{author}{\bibfnamefont{S.~A.} \bibnamefont{Yang}},
  \bibinfo{journal}{Phys. Rev. B} \textbf{\bibinfo{volume}{97}},
  \bibinfo{pages}{115125} (\bibinfo{year}{2018}).

\bibitem[{\citenamefont{Liang et~al.}(2016)\citenamefont{Liang, Zhou, Yu, Wang,
  and Weng}}]{PhysRevB.93.085427}
\bibinfo{author}{\bibfnamefont{Q.~F.} \bibnamefont{Liang}},
  \bibinfo{author}{\bibfnamefont{J.}~\bibnamefont{Zhou}},
  \bibinfo{author}{\bibfnamefont{R.}~\bibnamefont{Yu}},
  \bibinfo{author}{\bibfnamefont{Z.}~\bibnamefont{Wang}}, \bibnamefont{and}
  \bibinfo{author}{\bibfnamefont{H.}~\bibnamefont{Weng}},
  \bibinfo{journal}{Phys. Rev. B} \textbf{\bibinfo{volume}{93}},
  \bibinfo{pages}{085427} (\bibinfo{year}{2016}).

\bibitem[{\citenamefont{Zhang and Zhou}(2020)}]{PhysRevB.101.085202}
\bibinfo{author}{\bibfnamefont{S.B.} \bibnamefont{Zhang}} \bibnamefont{and}
  \bibinfo{author}{\bibfnamefont{J.}~\bibnamefont{Zhou}},
  \bibinfo{journal}{Phys. Rev. B} \textbf{\bibinfo{volume}{101}},
  \bibinfo{pages}{085202} (\bibinfo{year}{2020}).

\bibitem[{\citenamefont{Wang et~al.}(2015)\citenamefont{Wang, Lu, and
  Bertoldi}}]{PhysRevLett.115.104302}
\bibinfo{author}{\bibfnamefont{P.}~\bibnamefont{Wang}},
  \bibinfo{author}{\bibfnamefont{L.}~\bibnamefont{Lu}}, \bibnamefont{and}
  \bibinfo{author}{\bibfnamefont{K.}~\bibnamefont{Bertoldi}},
  \bibinfo{journal}{Phys. Rev. Lett.} \textbf{\bibinfo{volume}{115}},
  \bibinfo{pages}{104302} (\bibinfo{year}{2015}).

\bibitem[{\citenamefont{Xie et~al.}(2019)\citenamefont{Xie, Li, Ullah, Li,
  Wang, Li, Li, Yunoki, and Chen}}]{PhysRevB.99.174306}
\bibinfo{author}{\bibfnamefont{Q.}~\bibnamefont{Xie}},
  \bibinfo{author}{\bibfnamefont{J.}~\bibnamefont{Li}},
  \bibinfo{author}{\bibfnamefont{S.}~\bibnamefont{Ullah}},
  \bibinfo{author}{\bibfnamefont{R.}~\bibnamefont{Li}},
  \bibinfo{author}{\bibfnamefont{L.}~\bibnamefont{Wang}},
  \bibinfo{author}{\bibfnamefont{D.}~\bibnamefont{Li}},
  \bibinfo{author}{\bibfnamefont{Y.}~\bibnamefont{Li}},
  \bibinfo{author}{\bibfnamefont{S.}~\bibnamefont{Yunoki}}, \bibnamefont{and}
  \bibinfo{author}{\bibfnamefont{X.~Q.} \bibnamefont{Chen}},
  \bibinfo{journal}{Phys. Rev. B} \textbf{\bibinfo{volume}{99}},
  \bibinfo{pages}{174306} (\bibinfo{year}{2019}).

\bibitem[{\citenamefont{Liu et~al.}(2017{\natexlab{a}})\citenamefont{Liu, Xu,
  Zhang, and Duan}}]{PhysRevB.96.064106}
\bibinfo{author}{\bibfnamefont{Y.}~\bibnamefont{Liu}},
  \bibinfo{author}{\bibfnamefont{Y.}~\bibnamefont{Xu}},
  \bibinfo{author}{\bibfnamefont{S.~C.} \bibnamefont{Zhang}}, \bibnamefont{and}
  \bibinfo{author}{\bibfnamefont{W.}~\bibnamefont{Duan}},
  \bibinfo{journal}{Phys. Rev. B} \textbf{\bibinfo{volume}{96}},
  \bibinfo{pages}{064106} (\bibinfo{year}{2017}{\natexlab{a}}).

\bibitem[{\citenamefont{Jin et~al.}(2018)\citenamefont{Jin, Chen, Xia, Zhao,
  Wang, and Xu}}]{PhysRevB.98.220103}
\bibinfo{author}{\bibfnamefont{Y.~J.} \bibnamefont{Jin}},
  \bibinfo{author}{\bibfnamefont{Z.~J.} \bibnamefont{Chen}},
  \bibinfo{author}{\bibfnamefont{B.~W.} \bibnamefont{Xia}},
  \bibinfo{author}{\bibfnamefont{Y.~J.} \bibnamefont{Zhao}},
  \bibinfo{author}{\bibfnamefont{R.}~\bibnamefont{Wang}}, \bibnamefont{and}
  \bibinfo{author}{\bibfnamefont{H.}~\bibnamefont{Xu}}, \bibinfo{journal}{Phys.
  Rev. B} \textbf{\bibinfo{volume}{98}}, \bibinfo{pages}{220103}
  (\bibinfo{year}{2018}).

\bibitem[{\citenamefont{S{\"u}sstrunk and
  Huber}(2015{\natexlab{a}})}]{Susstrunk47}
\bibinfo{author}{\bibfnamefont{R.}~\bibnamefont{S{\"u}sstrunk}}
  \bibnamefont{and} \bibinfo{author}{\bibfnamefont{S.~D.} \bibnamefont{Huber}},
  \bibinfo{journal}{Science} \textbf{\bibinfo{volume}{349}},
  \bibinfo{pages}{47} (\bibinfo{year}{2015}{\natexlab{a}}).

\bibitem[{\citenamefont{Liu et~al.}(2018{\natexlab{a}})\citenamefont{Liu, Deng,
  and Wakabayashi}}]{PhysRevB.97.035442}
\bibinfo{author}{\bibfnamefont{F.}~\bibnamefont{Liu}},
  \bibinfo{author}{\bibfnamefont{H.~Y.} \bibnamefont{Deng}}, \bibnamefont{and}
  \bibinfo{author}{\bibfnamefont{K.}~\bibnamefont{Wakabayashi}},
  \bibinfo{journal}{Phys. Rev. B} \textbf{\bibinfo{volume}{97}},
  \bibinfo{pages}{035442} (\bibinfo{year}{2018}{\natexlab{a}}).

\bibitem[{\citenamefont{Stenull et~al.}(2016)\citenamefont{Stenull, Kane, and
  Lubensky}}]{PhysRevLett.117.068001}
\bibinfo{author}{\bibfnamefont{O.}~\bibnamefont{Stenull}},
  \bibinfo{author}{\bibfnamefont{C.~L.} \bibnamefont{Kane}}, \bibnamefont{and}
  \bibinfo{author}{\bibfnamefont{T.~C.} \bibnamefont{Lubensky}},
  \bibinfo{journal}{Phys. Rev. Lett.} \textbf{\bibinfo{volume}{117}},
  \bibinfo{pages}{068001} (\bibinfo{year}{2016}).



\bibitem[{\citenamefont{Liu et~al.}(2018{\natexlab{b}})\citenamefont{Liu, Xu,
  and Duan}}]{liu2018berry}
\bibinfo{author}{\bibfnamefont{Y.}~\bibnamefont{Liu}},
  \bibinfo{author}{\bibfnamefont{Y.}~\bibnamefont{Xu}}, \bibnamefont{and}
  \bibinfo{author}{\bibfnamefont{W.}~\bibnamefont{Duan}},
  \bibinfo{journal}{National Science Review} \textbf{\bibinfo{volume}{5}},
  \bibinfo{pages}{314} (\bibinfo{year}{2018}{\natexlab{b}}).

\bibitem[{\citenamefont{S{\"u}sstrunk and
  Huber}(2015{\natexlab{b}})}]{susstrunk2015observation}
\bibinfo{author}{\bibfnamefont{R.}~\bibnamefont{S{\"u}sstrunk}}
  \bibnamefont{and} \bibinfo{author}{\bibfnamefont{S.~D.} \bibnamefont{Huber}},
  \bibinfo{journal}{Science} \textbf{\bibinfo{volume}{349}},
  \bibinfo{pages}{47} (\bibinfo{year}{2015}{\natexlab{b}}).

\bibitem[{\citenamefont{Mousavi et~al.}(2015)\citenamefont{Mousavi, Khanikaev,
  and Wang}}]{mousavi2015topologically}
\bibinfo{author}{\bibfnamefont{S.~H.} \bibnamefont{Mousavi}},
  \bibinfo{author}{\bibfnamefont{A.~B.} \bibnamefont{Khanikaev}},
  \bibnamefont{and} \bibinfo{author}{\bibfnamefont{Z.}~\bibnamefont{Wang}},
  \bibinfo{journal}{Nature communications} \textbf{\bibinfo{volume}{6}},
  \bibinfo{pages}{1} (\bibinfo{year}{2015}).

\bibitem[{\citenamefont{He et~al.}(2016)\citenamefont{He, Ni, Ge, Sun, Chen,
  Lu, Liu, and Chen}}]{he2016acoustic}
\bibinfo{author}{\bibfnamefont{C.}~\bibnamefont{He}},
  \bibinfo{author}{\bibfnamefont{X.}~\bibnamefont{Ni}},
  \bibinfo{author}{\bibfnamefont{H.}~\bibnamefont{Ge}},
  \bibinfo{author}{\bibfnamefont{X.~C.} \bibnamefont{Sun}},
  \bibinfo{author}{\bibfnamefont{Y.~B.} \bibnamefont{Chen}},
  \bibinfo{author}{\bibfnamefont{M.~H.} \bibnamefont{Lu}},
  \bibinfo{author}{\bibfnamefont{X.~P.} \bibnamefont{Liu}}, \bibnamefont{and}
  \bibinfo{author}{\bibfnamefont{Y.~F.} \bibnamefont{Chen}},
  \bibinfo{journal}{Nature physics} \textbf{\bibinfo{volume}{12}},
  \bibinfo{pages}{1124} (\bibinfo{year}{2016}).

\bibitem[{\citenamefont{Zhang et~al.}(2019)\citenamefont{Zhang, Miao, Wang,
  Lin, Cao, Fabbris, Said, Liu, Lei, Fang et~al.}}]{PhysRevLett.123.245302}
\bibinfo{author}{\bibfnamefont{T.~T.} \bibnamefont{Zhang}},
  \bibinfo{author}{\bibfnamefont{H.}~\bibnamefont{Miao}},
  \bibinfo{author}{\bibfnamefont{Q.}~\bibnamefont{Wang}},
  \bibinfo{author}{\bibfnamefont{J.~Q.} \bibnamefont{Lin}},
  \bibinfo{author}{\bibfnamefont{Y.}~\bibnamefont{Cao}},
  \bibinfo{author}{\bibfnamefont{G.}~\bibnamefont{Fabbris}},
  \bibinfo{author}{\bibfnamefont{A.~H.} \bibnamefont{Said}},
  \bibinfo{author}{\bibfnamefont{X.}~\bibnamefont{Liu}},
  \bibinfo{author}{\bibfnamefont{H.~C.} \bibnamefont{Lei}},
  \bibinfo{author}{\bibfnamefont{Z.}~\bibnamefont{Fang}}, \bibnamefont{et~al.},
  \bibinfo{journal}{Phys. Rev. Lett.} \textbf{\bibinfo{volume}{123}},
  \bibinfo{pages}{245302} (\bibinfo{year}{2019}).

\bibitem[{\citenamefont{Peng et~al.}(2020)\citenamefont{Peng, Hu, Murakami,
  Zhang, and Monserrat}}]{peng2020topological}
\bibinfo{author}{\bibfnamefont{B.}~\bibnamefont{Peng}},
  \bibinfo{author}{\bibfnamefont{Y.}~\bibnamefont{Hu}},
  \bibinfo{author}{\bibfnamefont{S.}~\bibnamefont{Murakami}},
  \bibinfo{author}{\bibfnamefont{T.}~\bibnamefont{Zhang}}, \bibnamefont{and}
  \bibinfo{author}{\bibfnamefont{B.}~\bibnamefont{Monserrat}},
  \bibinfo{journal}{Science advances} \textbf{\bibinfo{volume}{6}},
  \bibinfo{pages}{eabd1618} (\bibinfo{year}{2020}).

\bibitem[{\citenamefont{Li et~al.}(2020)\citenamefont{Li, Xie, Liu, Li, Liu,
  Wang, Li, Li, and Chen}}]{PhysRevB.101.024301}
\bibinfo{author}{\bibfnamefont{J.}~\bibnamefont{Li}},
  \bibinfo{author}{\bibfnamefont{Q.}~\bibnamefont{Xie}},
  \bibinfo{author}{\bibfnamefont{J.}~\bibnamefont{Liu}},
  \bibinfo{author}{\bibfnamefont{R.}~\bibnamefont{Li}},
  \bibinfo{author}{\bibfnamefont{M.}~\bibnamefont{Liu}},
  \bibinfo{author}{\bibfnamefont{L.}~\bibnamefont{Wang}},
  \bibinfo{author}{\bibfnamefont{D.}~\bibnamefont{Li}},
  \bibinfo{author}{\bibfnamefont{Y.}~\bibnamefont{Li}}, \bibnamefont{and}
  \bibinfo{author}{\bibfnamefont{X.~Q.} \bibnamefont{Chen}},
  \bibinfo{journal}{Phys. Rev. B} \textbf{\bibinfo{volume}{101}},
  \bibinfo{pages}{024301} (\bibinfo{year}{2020}).

\bibitem[{\citenamefont{See Supplemental Material for the~computational method
  and effective~model analysis}()}]{SM}
 \bibnamefont{See Supplemental Material  for the
computational method, the detailed symmetry and effective
model analysis, the summary of space groups, which
includes Refs. [35-44]}.

\bibitem[{\citenamefont{Berry}(1984)}]{rspa.1984.0023}
\bibinfo{author}{\bibfnamefont{M.~V.} \bibnamefont{Berry}},
  \bibinfo{journal}{Proceedings of the Royal Society of London. A. Mathematical
  and Physical Sciences} \textbf{\bibinfo{volume}{392}}, \bibinfo{pages}{45}
  (\bibinfo{year}{1984}).

\bibitem[{\citenamefont{Kohn and Sham}(1965)}]{kohn1965self}
\bibinfo{author}{\bibfnamefont{W.}~\bibnamefont{Kohn}} \bibnamefont{and}
  \bibinfo{author}{\bibfnamefont{L.~J.} \bibnamefont{Sham}},
  \bibinfo{journal}{Physical review} \textbf{\bibinfo{volume}{140}},
  \bibinfo{pages}{A1133} (\bibinfo{year}{1965}).

\bibitem[{\citenamefont{Kresse and Furthm\"uller}(1996)}]{PhysRevB.54.11169}
\bibinfo{author}{\bibfnamefont{G.}~\bibnamefont{Kresse}} \bibnamefont{and}
  \bibinfo{author}{\bibfnamefont{J.}~\bibnamefont{Furthm\"uller}},
  \bibinfo{journal}{Phys. Rev. B} \textbf{\bibinfo{volume}{54}},
  \bibinfo{pages}{11169} (\bibinfo{year}{1996}).

\bibitem[{\citenamefont{Kresse and
  Furthm{\"u}ller}(1996)}]{kresse1996efficiency}
\bibinfo{author}{\bibfnamefont{G.}~\bibnamefont{Kresse}} \bibnamefont{and}
  \bibinfo{author}{\bibfnamefont{J.}~\bibnamefont{Furthm{\"u}ller}},
  \bibinfo{journal}{Computational materials science}
  \textbf{\bibinfo{volume}{6}}, \bibinfo{pages}{15} (\bibinfo{year}{1996}).

\bibitem[{\citenamefont{Perdew et~al.}(1996)\citenamefont{Perdew, Burke, and
  Ernzerhof}}]{PhysRevLett.77.3865}
\bibinfo{author}{\bibfnamefont{J.~P.} \bibnamefont{Perdew}},
  \bibinfo{author}{\bibfnamefont{K.}~\bibnamefont{Burke}}, \bibnamefont{and}
  \bibinfo{author}{\bibfnamefont{M.}~\bibnamefont{Ernzerhof}},
  \bibinfo{journal}{Phys. Rev. Lett.} \textbf{\bibinfo{volume}{77}},
  \bibinfo{pages}{3865} (\bibinfo{year}{1996}).

\bibitem[{\citenamefont{Perdew et~al.}(1997)\citenamefont{Perdew, Burke, and
  Ernzerhof}}]{PhysRevLett.78.1396}
\bibinfo{author}{\bibfnamefont{J.~P.} \bibnamefont{Perdew}},
  \bibinfo{author}{\bibfnamefont{K.}~\bibnamefont{Burke}}, \bibnamefont{and}
  \bibinfo{author}{\bibfnamefont{M.}~\bibnamefont{Ernzerhof}},
  \bibinfo{journal}{Phys. Rev. Lett.} \textbf{\bibinfo{volume}{78}},
  \bibinfo{pages}{1396} (\bibinfo{year}{1997}).

\bibitem[{\citenamefont{Kresse and Joubert}(1999)}]{PhysRevB.59.1758}
\bibinfo{author}{\bibfnamefont{G.}~\bibnamefont{Kresse}} \bibnamefont{and}
  \bibinfo{author}{\bibfnamefont{D.}~\bibnamefont{Joubert}},
  \bibinfo{journal}{Phys. Rev. B} \textbf{\bibinfo{volume}{59}},
  \bibinfo{pages}{1758} (\bibinfo{year}{1999}).

\bibitem[{\citenamefont{Ceperley and Alder}(1980)}]{PhysRevLett.45.566}
\bibinfo{author}{\bibfnamefont{D.~M.} \bibnamefont{Ceperley}} \bibnamefont{and}
  \bibinfo{author}{\bibfnamefont{B.~J.} \bibnamefont{Alder}},
  \bibinfo{journal}{Phys. Rev. Lett.} \textbf{\bibinfo{volume}{45}},
  \bibinfo{pages}{566} (\bibinfo{year}{1980}).

\bibitem[{\citenamefont{Togo and Tanaka}(2015)}]{TOGO20151}
\bibinfo{author}{\bibfnamefont{A.}~\bibnamefont{Togo}} \bibnamefont{and}
  \bibinfo{author}{\bibfnamefont{I.}~\bibnamefont{Tanaka}},
  \bibinfo{journal}{Scripta Materialia} \textbf{\bibinfo{volume}{108}},
  \bibinfo{pages}{1 } (\bibinfo{year}{2015}).

\bibitem[{\citenamefont{Sancho et~al.}(1984)\citenamefont{Sancho, Sancho, and
  Rubio}}]{Sancho_1984}
\bibinfo{author}{\bibfnamefont{M.~P.~L.} \bibnamefont{Sancho}},
  \bibinfo{author}{\bibfnamefont{J.~M.~L.} \bibnamefont{Sancho}},
  \bibnamefont{and} \bibinfo{author}{\bibfnamefont{J.}~\bibnamefont{Rubio}},
  \bibinfo{journal}{Journal of Physics F: Metal Physics}
  \textbf{\bibinfo{volume}{14}}, \bibinfo{pages}{1205} (\bibinfo{year}{1984}).

\bibitem[{\citenamefont{Wu et~al.}(2018)\citenamefont{Wu, Zhang, Song, Troyer,
  and Soluyanov}}]{WU2018405}
\bibinfo{author}{\bibfnamefont{Q.}~\bibnamefont{Wu}},
  \bibinfo{author}{\bibfnamefont{S.}~\bibnamefont{Zhang}},
  \bibinfo{author}{\bibfnamefont{H.~F.} \bibnamefont{Song}},
  \bibinfo{author}{\bibfnamefont{M.}~\bibnamefont{Troyer}}, \bibnamefont{and}
  \bibinfo{author}{\bibfnamefont{A.~A.} \bibnamefont{Soluyanov}},
  \bibinfo{journal}{Computer Physics Communications}
  \textbf{\bibinfo{volume}{224}}, \bibinfo{pages}{405 } (\bibinfo{year}{2018}).



\end{thebibliography}

\begin{thebibliography}{10}
\expandafter\ifx\csname natexlab\endcsname\relax\def\natexlab#1{#1}\fi
\expandafter\ifx\csname bibnamefont\endcsname\relax
  \def\bibnamefont#1{#1}\fi
\expandafter\ifx\csname bibfnamefont\endcsname\relax
  \def\bibfnamefont#1{#1}\fi
\expandafter\ifx\csname citenamefont\endcsname\relax
  \def\citenamefont#1{#1}\fi
\expandafter\ifx\csname url\endcsname\relax
  \def\url#1{\texttt{#1}}\fi
\expandafter\ifx\csname urlprefix\endcsname\relax\def\urlprefix{URL }\fi
\providecommand{\bibinfo}[2]{#2}
\providecommand{\eprint}[2][]{\url{#2}}

\bibitem[{\citenamefont{Kohn and Sham}(1965)}]{kohn1965self}
\bibinfo{author}{\bibfnamefont{W.}~\bibnamefont{Kohn}} \bibnamefont{and}
  \bibinfo{author}{\bibfnamefont{L.~J.} \bibnamefont{Sham}},
  \bibinfo{journal}{Physical review} \textbf{\bibinfo{volume}{140}},
  \bibinfo{pages}{A1133} (\bibinfo{year}{1965}).

\bibitem[{\citenamefont{Kresse and Furthm\"uller}(1996)}]{PhysRevB.54.11169}
\bibinfo{author}{\bibfnamefont{G.}~\bibnamefont{Kresse}} \bibnamefont{and}
  \bibinfo{author}{\bibfnamefont{J.}~\bibnamefont{Furthm\"uller}},
  \bibinfo{journal}{Phys. Rev. B} \textbf{\bibinfo{volume}{54}},
  \bibinfo{pages}{11169} (\bibinfo{year}{1996}).

\bibitem[{\citenamefont{Kresse and
  Furthm{\"u}ller}(1996)}]{kresse1996efficiency}
\bibinfo{author}{\bibfnamefont{G.}~\bibnamefont{Kresse}} \bibnamefont{and}
  \bibinfo{author}{\bibfnamefont{J.}~\bibnamefont{Furthm{\"u}ller}},
  \bibinfo{journal}{Computational materials science}
  \textbf{\bibinfo{volume}{6}}, \bibinfo{pages}{15} (\bibinfo{year}{1996}).

\bibitem[{\citenamefont{Perdew et~al.}(1996)\citenamefont{Perdew, Burke, and
  Ernzerhof}}]{PhysRevLett.77.3865}
\bibinfo{author}{\bibfnamefont{J.~P.} \bibnamefont{Perdew}},
  \bibinfo{author}{\bibfnamefont{K.}~\bibnamefont{Burke}}, \bibnamefont{and}
  \bibinfo{author}{\bibfnamefont{M.}~\bibnamefont{Ernzerhof}},
  \bibinfo{journal}{Phys. Rev. Lett.} \textbf{\bibinfo{volume}{77}},
  \bibinfo{pages}{3865} (\bibinfo{year}{1996}).

\bibitem[{\citenamefont{Perdew et~al.}(1997)\citenamefont{Perdew, Burke, and
  Ernzerhof}}]{PhysRevLett.78.1396}
\bibinfo{author}{\bibfnamefont{J.~P.} \bibnamefont{Perdew}},
  \bibinfo{author}{\bibfnamefont{K.}~\bibnamefont{Burke}}, \bibnamefont{and}
  \bibinfo{author}{\bibfnamefont{M.}~\bibnamefont{Ernzerhof}},
  \bibinfo{journal}{Phys. Rev. Lett.} \textbf{\bibinfo{volume}{78}},
  \bibinfo{pages}{1396} (\bibinfo{year}{1997}).

\bibitem[{\citenamefont{Kresse and Joubert}(1999)}]{PhysRevB.59.1758}
\bibinfo{author}{\bibfnamefont{G.}~\bibnamefont{Kresse}} \bibnamefont{and}
  \bibinfo{author}{\bibfnamefont{D.}~\bibnamefont{Joubert}},
  \bibinfo{journal}{Phys. Rev. B} \textbf{\bibinfo{volume}{59}},
  \bibinfo{pages}{1758} (\bibinfo{year}{1999}).

\bibitem[{\citenamefont{Ceperley and Alder}(1980)}]{PhysRevLett.45.566}
\bibinfo{author}{\bibfnamefont{D.~M.} \bibnamefont{Ceperley}} \bibnamefont{and}
  \bibinfo{author}{\bibfnamefont{B.~J.} \bibnamefont{Alder}},
  \bibinfo{journal}{Phys. Rev. Lett.} \textbf{\bibinfo{volume}{45}},
  \bibinfo{pages}{566} (\bibinfo{year}{1980}).

\bibitem[{\citenamefont{Togo and Tanaka}(2015)}]{TOGO20151}
\bibinfo{author}{\bibfnamefont{A.}~\bibnamefont{Togo}} \bibnamefont{and}
  \bibinfo{author}{\bibfnamefont{I.}~\bibnamefont{Tanaka}},
  \bibinfo{journal}{Scripta Materialia} \textbf{\bibinfo{volume}{108}},
  \bibinfo{pages}{1 } (\bibinfo{year}{2015}).

\bibitem[{\citenamefont{Sancho et~al.}(1984)\citenamefont{Sancho, Sancho, and
  Rubio}}]{Sancho_1984}
\bibinfo{author}{\bibfnamefont{M.~P.~L.} \bibnamefont{Sancho}},
  \bibinfo{author}{\bibfnamefont{J.~M.~L.} \bibnamefont{Sancho}},
  \bibnamefont{and} \bibinfo{author}{\bibfnamefont{J.}~\bibnamefont{Rubio}},
  \bibinfo{journal}{Journal of Physics F: Metal Physics}
  \textbf{\bibinfo{volume}{14}}, \bibinfo{pages}{1205} (\bibinfo{year}{1984}).

\bibitem[{\citenamefont{Wu et~al.}(2018)\citenamefont{Wu, Zhang, Song, Troyer,
  and Soluyanov}}]{WU2018405}
\bibinfo{author}{\bibfnamefont{Q.}~\bibnamefont{Wu}},
  \bibinfo{author}{\bibfnamefont{S.}~\bibnamefont{Zhang}},
  \bibinfo{author}{\bibfnamefont{H.-F.} \bibnamefont{Song}},
  \bibinfo{author}{\bibfnamefont{M.}~\bibnamefont{Troyer}}, \bibnamefont{and}
  \bibinfo{author}{\bibfnamefont{A.~A.} \bibnamefont{Soluyanov}},
  \bibinfo{journal}{Computer Physics Communications}
  \textbf{\bibinfo{volume}{224}}, \bibinfo{pages}{405 } (\bibinfo{year}{2018}).

\end{thebibliography}
\end{document}